\documentclass[12pt]{article}
\usepackage{graphicx,geometry,indentfirst,setspace,booktabs,array,caption,amsmath,placeins,tabularx,titling,lipsum}
\usepackage[style=apa]{biblatex}
\usepackage[labelfont=bf,skip=12pt]{caption}
\graphicspath{{Figures/}}

\newcolumntype{L}[1]{>{\raggedright\arraybackslash}p{#1}}
\newcolumntype{C}[1]{>{\centering\arraybackslash}p{#1}}
\newcolumntype{R}[1]{>{\raggedleft\arraybackslash}p{#1}}
\newcolumntype{W}{>{\raggedright\arraybackslash}X}
\newcolumntype{Y}{>{\centering\arraybackslash}X}
\newcolumntype{Z}{>{\raggedleft\arraybackslash}X}

\begin{document}

\begin{titlepage}

\renewcommand{\thefootnote}{}
\footnotetext{The author is grateful to Nicholas Buchholz and Brian Weeden for their assistance in crafting this paper.}
\renewcommand{\thefootnote}{\arabic{footnote}}  

\title{\large\bfseries VALUING DIFFUSE GLOBAL PUBLIC GOODS FROM SATELLITE CONSTELLATIONS: EVIDENCE FROM GPS AND AIRLINE DELAYS}
\author{Lev Ricanati\thanks{Department of Economics, Princeton University, levr@princeton.edu.}}
\maketitle

\vspace{-1.5em}
\begin{abstract}
\noindent This paper studies the welfare impact of discrete improvements to global public goods in the context of the Global Positioning System (GPS). Specifically, I find that by disabling Selective Availability in May, 2000, and thus significantly increasing the accuracy of GPS, the United States generated at least \$268 million (2000 dollars) of additional welfare gains. To quantify this welfare impact, I apply a difference-in-differences model to the Bureau of Transportation Statistics’s Airline On-Time Performance Data in the years 1999 and 2000. I use this model to estimate the time saved per flight attributable to the improved GPS and multiply these time savings by the number of passengers in the ensuing year and their values of time. I conclude by estimating the economic loss from current threats to the provision of satellite-based global public goods.

\end{abstract}

\setcounter{page}{0}
\thispagestyle{empty}
\end{titlepage}

\newpage
\tableofcontents

\newpage
\section{Introduction}
Due to the immense welfare benefits generated by global public goods (GPGs), small improvements in their efficiency can lead to large increases in global welfare. Classic examples of GPGs include global standard regulatory practices, weather monitoring satellites, and the preservation of diverse ecosystems. The welfare benefits of GPGs are often quite diffuse meaning that benefits are large in aggregate rather than per individual. One important GPG with diffuse benefits is the Global Positioning System (GPS). GPS is a network of 31 satellites owned and operated by the United States (US) that provides anyone around the globe with their location free of charge \parencite{noauthor_gpsgov_nodate-1}.

In this paper, I investigate the impact of discrete improvements to satellite based GPGs, specifically by analyzing an accuracy improvement to GPS in May 2000. I find that the improvement led to a welfare gain of \$268 million (2000 dollars) in the year following implementation.\footnote{Equivalent to just over \$504 million in 2025 dollars \parencite{noauthor_cpi_nodate}.} This amount is more than twice what the US congress appropriated for the system in 2000 \parencite{noauthor_h_1999}. Importantly, the welfare gains I calculate are only a small part of the total gains from this improvement. The increased accuracy of GPS improved the commercial viability of existing GPS-reliant industries and spawned entirely new ones such as smartphone navigation and precision agriculture \parencite{oconnor_economic_2019}. The total welfare generated by this improvement, both from its impact on existing industries and its enabling of new ones, was likely in the billions of dollars each year. This analysis illustrates how GPGs turn efficiency improvements into massive welfare gains. This is particularly important to understand amidst the recent rise in the use of GPS jamming and spoofing technologies in conflict zones, which has disrupted commercial flights and erased the benefits wrought from GPS.  

When it was initially designed and built, the widely available civil GPS signal was intentionally degraded in a policy known as Selective Availability (SA) which excluded non-military users from accessing precise GPS. While Selective Availability was in place, the civil GPS signal was only accurate to 300 feet, equivalent to the difference of trying to land on a runway and ending up one or two runway widths away \parencite{noauthor_gpsgov_nodate}. At midnight on May 2nd, 2000, President Clinton disabled Selective Availability, instantly making the civil signal 10 times more accurate \parencite{noauthor_brief_2024}. This decision transformed GPS overnight into a usable global public good without any cost incurred by the government or consumers.

I focus my analysis of the economic effects of this decision on US airline delays. Using a government dataset of all domestic flights and their delays, I run a difference-in-differences regression to isolate the effect of SA on flight delays. Due to the lack of a natural control, I split the data and designate flights in 1999 as the control group and flights in 2000 as the treatment group. I conduct a series of regressions segmented by flight distance and amount of time post-treatment. The results of my analysis show that the removal of SA led to a general reduction in airline delays of 0.617 minutes per flight with savings of up to 3 minutes on longer flights in the first few months. I hypothesize that the mechanism for the reduced delays was pilots’ ability to adhere to flight plans more accurately throughout their flights.

To estimate the welfare gains, I multiply the time saved per flight by the number of domestic US passengers per month and the Department of Transportation’s estimates for how much passengers value the time they spend traveling by air. This method results in a value of \$268 million in welfare savings from all US passengers in the year following May 2nd, 2000.

Global satellite constellations, like post-SA GPS, are true global public goods. They provide non-rival and non-excludable benefits to anyone with a receiver, which are relatively cheap to acquire. Welfare benefits from these constellations are accrued around the globe and are therefore difficult to quantify. This becomes a problem for governments deciding whether or not to invest in these systems. Lacking an easy cost-benefit analysis makes it more difficult to argue for the provision or improvement of satellite based GPGs. This can slow down the implementation of welfare-enhancing upgrades. This paper demonstrates that the removal of SA was indeed welfare-enhancing, hopefully motivating future investments into protecting and improving GPS services.

I start by discussing relevant literature and then provide a brief history of GPS and SA. I proceed to talk about the origins of the data, including summary statistics and graphs. I then discuss the methodology I use as well as the assumptions I make, followed by the regression results. The results include a preliminary quantification of the welfare losses from 2 current threats to GPS provision, jamming and spoofing. Finally, I present policy recommendations for ensuring continuous GPS as well as directions for future research.

\subsection{Literature Review}
A lot has been written about the economic impact of GPS, both on existing markets and through the creation of novel ones. Two oft cited reports, one in 2011 and the other in 2019, attempt to quantify the economic benefits of GPS \parencite{oconnor_economic_2019,pham_economic_2011}. The 2011 report places the economic benefit at \$67.6 billion per year in the US. The author recognizes the growing nature of GPS enabled industries and predicts that this economic benefit could reach \$122.4 billion per year \parencite[p.~1]{pham_economic_2011}. The later report, in 2019, concludes that GPS added \$1.4 trillion in economic benefits since it was opened to public use in 1983 with about 90\% of those benefits accumulating in the 2010s \parencite[p.~ES-3]{oconnor_economic_2019}. These reports also attempt to quantify what the cost of a disruption to GPS would be. \textcite{pham_economic_2011} calculates the economic loss from a GPS disruption to be equal to the current benefit of GPS plus the monetary value invested into GPS technology, such as receivers \parencite[p.~12]{pham_economic_2011}. One problem with this approach is that some industries would be able to revert to other systems or practices in the event of an outage. These industries would still then be able to generate some economic activity, just not at the levels that GPS enabled. \textcite{oconnor_economic_2019} take this into consideration in the methodology of their report leading to an estimation of \$1 billion lost per day in the event of a GPS outage \parencite[p.~ES-4]{oconnor_economic_2019}.\footnote{They assume that other countries' satellite navigation systems are also disabled.} As the authors acknowledge, this number is probably an underestimate because they did not include industries where GPS has led to efficiency gains but was not central to the performance of the industry. One of the industries they chose not to study was aviation, leaving a hole in the literature on the economic benefits of GPS.

There has also been very little study of the Selective Availability decision and its economic impact. The literature that does exist on SA focuses more on the political decision making surrounding its removal rather than the economic effects \parencite{gleason_galileo_2009, weeden_case_2017}. One unpublished cost-benefit analysis of the decision concluded that “the clear decision from an efficiency standpoint [was] to eliminate SA”, however, it stopped short of quantifying this efficiency gain \parencite[p.~24]{weeden_is_2013}. Likewise, a 2002 paper charting the evolution of GPS, predicted that GPS, recently at its full commercial potential, could reduce flight times by 2\% due to efficiencies in routing planes \parencite[p.~68]{kumar_evolution_2002}. My paper intends to add to the scholarship on the economic impact of GPS by addressing the lack of quantification of the economic value to the US aviation industry generated by removing SA. My decision to focus on the impact on aviation is motivated by both the lack of previous literature on the specific effects of GPS on aviation and the use of improved aviation performance as one of the main arguments made for turning off SA \parencite[p.~151]{weeden_case_2017}.

In quantifying the value of turning off SA, I utilize concepts from the economics literature on how consumers value their time. The economics field of transportation analysis contains several analyses of consumers’ Value of Time (VOT), i.e. how much a consumer would be willing to pay for a reduction in travel time. VOT numbers are influenced by factors such as the reliability of the travel, costs incurred, and the total travel time \parencite{small_valuation_2012}. VOT literature has focused on different components of the travel experience. \textcite{small_valuation_2012} distinguishes between analyses that focus on consumer Value of Reliability (VOR) and VOT, referring primarily to the time spent in transit \parencite[p.~2]{small_valuation_2012}. Another approach to VOT is to quantify consumer willingness to pay for reductions in pre-travel waiting times. This is useful for evaluating technologies such as ride-hailing apps that attempt to make waiting times transparent and shorter \parencite{buchholz_personalized_2024}. The analysis in this paper centers on a technology, GPS, that reduced inflight delays compared to pre-GPS flights. Passengers derived benefits from these reduced delays through having more time post-flight to allocate to work or leisure. In order to quantify the value of this additional time, I utilize the Department of Transportation’s (DOT) estimation of air travelers’ valuations of reduced travel times. 

Finally, my paper is situated within the literature on global public goods. Classic GPG literature defines them as public goods that affect a majority of countries and that are provisioned by multiple countries \parencite{buchholz_global_2021}. Examples include combating climate change, setting regulatory standards, or fighting a pandemic. Under this definition, GPS is a different kind of GPG because it can benefit anyone on the globe but is only provisioned and paid for by one country. This is partially because it was originally developed as a military system for global military operations, not as a public good for commercial use. GPS’s development and eventual conversion into a GPG can be explained using the framework developed by \textcite{boadway_country_1999} for state decision making related to GPGs. They posit that states weigh the benefit to their citizens in addition to the benefit to the world when deciding whether to contribute to a GPG. Therefore, the US would want to develop GPS because of the large benefit it could provide to its citizens, both economically and as part of the public good of military protection.

\subsection{Background}
GPS-enabled navigation is a relatively simple system. At all times, a constellation of, at minimum, 24 GPS satellites are orbiting the Earth such that any user always has 4 satellites overhead. Each satellite sends out radio signals with their location and the time of the signal. Using this information, GPS receivers triangulate users’ positions in 3 dimensions. 

The first modern GPS satellite was launched in 1978 by the Department of Defense (DOD). Originally, the system was designed purely for military use and the location and timing signals were encrypted. GPS was indeed quite important to the military and contributed to the success of US ground forces in the 1991 Gulf War \parencite{noauthor_evolution_2016}. However, in 1983, a passenger airliner was shot down after drifting into Soviet airspace prompting President Reagan to decree that GPS should be available to the public for free \parencite{pellerin_united_2013}. In order to maintain its military edge, the DOD implemented a policy of Selective Availability. SA involved intentionally introducing errors into the GPS signals such that non-military users were not able to pinpoint their location with a high degree of accuracy. As one US Navy official put it, it was the difference between “dropping a bomb in the stadium… [and] dropping a bomb in the huddle” \parencite[p.~151]{weeden_case_2017}. A debate emerged in the 1990s over whether or not to disable SA. One of the major points of contention in this debate was whether the commercial gain would be worth sacrificing the military advantage of excludable GPS \parencite{weeden_case_2017}. As one DOD employee said in 1996, “something that's always been lacking… is the ability to assign specific costs and benefits of the GPS contribution to society” \parencite[p.~155]{weeden_case_2017}. This debate went back and forth between the FAA and the DOD with little progress towards resolving it. 

However, on May 1st, 2000, President Clinton declared that he was disabling SA at midnight that night. This decision came as a surprise to most people because the accepted timeline was that SA would be disabled at some point in the mid-2000s \parencite[p.~186]{weeden_case_2017}. People familiar with the decision say that it was in part motivated by the European push to get their competing global navigation satellite system, Galileo, launched and operating \parencite[p.~187]{weeden_case_2017}. Galileo has since been deployed but the decision to turn off SA before any Galileo satellites could be launched very likely contributed to GPS becoming the dominant and standard global navigation satellite system. 

 Since the public was given access to GPS’s full accuracy, many commercial industries have embraced the technology. These include existing industries such as shipping and aviation and brand new industries such as on-demand ride hailing, smartphone navigation, and precision agriculture. One such application of GPS made possible by the removal of SA was the advent of Geocaching, a global community of players who hide containers and publish the coordinates online for others to find using GPS. Public excitement about the removal of SA is evident in the fact that the first Geocache was placed on May 3rd, 2000, one day after SA was disabled \parencite{brigitte_happy_2020}.

\section{Data}
The main data source for this paper is the Bureau of Transportation Statistics’s (BTS) Airline On-Time Performance Data \parencite{noauthor_airline_nodate}. This dataset includes routing and performance data for reported domestic US flights in a selected time period. The flights were reported to BTS by all airlines that were responsible for at least 1 percent of domestic scheduled passenger revenues. This paper specifically looks at all flights in the years 1999 and 2000. In total, there are 11,210,931 observations in the dataset split into 5,527,884 in 1999 and 5,683,047 in 2000.

The cleaned quantitative variables for each flight are summarized in Table~\ref{tab:sumstats}.\footnote{Data cleaning procedures are described in Appendix~\ref{sec:datacleaning}. In total, 631 flights were removed from the dataset for potential inaccuracies.} All of the variables were part of the original dataset except \emph{adddelay}.

\begin{table}[h!]
\caption{Dataset Summary Statistics}
\label{tab:sumstats}

\centering
\begin{tabular}{L{2.5cm} L{4cm} R{2cm} R{2cm} R{2cm} R{2cm}}
\toprule
\textbf{Variable} & \textbf{Description} & \textbf{Minimum} & \textbf{Maximum} & \textbf{Mean} & \textbf{SD} \\
\midrule
\textit{date} & Date of the flight & 01jan1999 & 31dec2000 & N/A & N/A \\
\addlinespace
\textit{dep\_delay} & The difference between actual and scheduled departure time (minutes) & -990 & 1740 & 10.24 & 31.28 \\
\addlinespace
\textit{arr\_delay} & The difference between actual and scheduled arrival time (minutes) & -989 & 1724 & 9.37 & 34.57 \\
\addlinespace
\textit{cancelled} & A dummy variable for if the flight was not flown & 0 & 1 & 0.03 & 0.17 \\
\addlinespace
\textit{actual\_time} & The actual time between gate departure and gate arrival (minutes) & 10 & 850 & 127.73 & 70.69 \\
\addlinespace
\textit{air\_time} & The total time a flight was in air (minutes) & 1 & 750 & 105.59 & 66.95 \\
\addlinespace
\textit{distance} & Distance between origin and destination airports (miles) & 11 & 4962 & 758.84 & 568.08 \\
\addlinespace
\textit{adddelay} & Change in delay from dep\_delay to arr\_delay (minutes) & -127 & 481 & -0.80 & 13.91 \\
\bottomrule

\end{tabular}
\end{table}

In addition to the quantitative variables, the dataset contained information on a given flight’s origin, destination, and airline. Due to the reporting requirements, the dataset contained flights from the 11 largest airlines at the time. These airlines are American Airlines (AA), Alaska Airlines (AS), Continental Airlines (CO), Delta Airlines (DL), America West Airlines (HP), Aloha Air Cargo (KH), Northwest Airlines (NW),  Trans World Airways (TW), United Air Lines (UA), US Helicopter Corporation (US), and Southwest Airlines (WN).

US Helicopter Corporation’s inclusion in the dataset helps explain the observations with low flight distances. The shortest flights were helicopter rides between regional airports such as John F Kennedy and LaGuardia in New York.

\subsection{Creation of Additional Delay Variable}
While the reported variables provide valuable insight on the flights in the dataset, they were not ideal for testing my hypothesis. For example, the distance variable only measures distance between airports, not distance traveled, and thus is not directly affected by technology improvements. Additionally, while the arrival and departure delay values changed over time as a result of GPS enhancement, they are predominantly affected by changing airport infrastructure, FAA regulations, and other external factors.

In order to test the effect of the GPS enhancement, I introduced a new variable for the additional in-flight delay, \textit{adddelay}. \textit{Adddelay} was calculated by subtracting departure delays from arrival delays. Thus, if a flight took longer than expected due to in-flight causes, it would be delayed more when it landed and \textit{adddelay} would be positive. Conversely, if a flight gained time in-flight relative to its itinerary, \textit{adddelay} would be negative. This new variable isolates the in-flight delays, which is where I expect the effects of removing SA to manifest. 

\begin{figure}[h!]
\caption{Daily In-flight Delays}
\label{fig:dailydelays2yrs}
\centering
\includegraphics[width=0.9\textwidth]{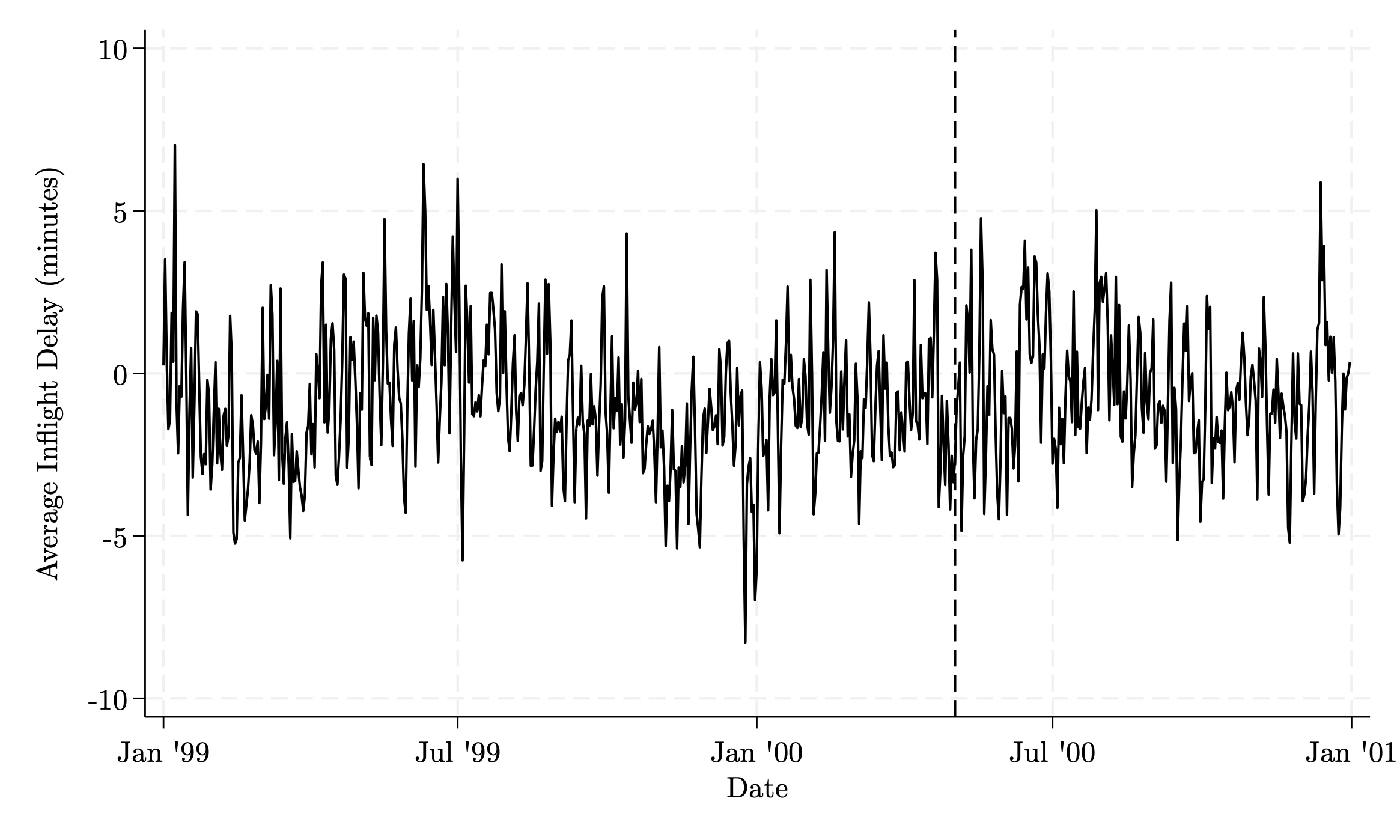}
\caption*{\small \textit{Note:} The dashed line is on May 2nd, 2000. SA was disabled from that day on.}
\end{figure}

An important aspect of the dataset that the \textit{adddelay} variable reveals is the seasonality of airline delays. Due to holidays, weather, or other reasons, the amount of airline travelers fluctuates over time. This leads to periods in which airports and the national air system need to handle large amounts of demand. Figure~\ref{fig:dailydelays2yrs} shows average daily \textit{adddelay} over the two years in the dataset. The dashed line indicates when SA was disabled.

\FloatBarrier

As can be seen in Figure~\ref{fig:dailydelays2yrs}, the variability in delays exists on the day by day and  seasonal levels. The daily peaks are most likely weekends, while the seasonal peaks occur in the summer months and around the December holidays. This seasonality helped motivate my choice of methodology as I discuss later. Interestingly, the graph appears to show more pronounced valleys and peaks before GPS SA was disabled. While there is not an easily identifiable drop in delays after GPS was improved, this smoothing of the graph points to the accuracy improvement having an effect on delays.

\subsection{Creation of Control Variables}
Control variables were designed in order to properly isolate the effect of disabling SA. While many factors might affect delays, most of them occur both in the control and treated groups of data, which my methodology is designed to account for. However, one main variable that has a large impact on delays and changes over time is inclement weather, such as rain, snow, or low visibility.

The dataset does not provide weather data for the flights. To address this problem, BTS introduced a requirement to report if the cause of the delay was weather. Unfortunately, the requirement was implemented in 2003, after the period of analysis. One method to control for weather disruptions is to acquire precipitation data from the National Oceanic and Atmospheric Administration for each departure and arrival airport. This approach would require integrating 2 years of data for over 100 airport locations.

Another option was to have cancellations serve as a proxy for poor weather and other airport disruptions. If the weather is bad enough, planes will not be able to land at destination airports nor takeoff from origin airports. This leads to an increase in cancellations which is reflected in the dataset. Thus a measure of the percent of flights canceled per day per airport should approximate whether the weather is poor or not at that airport. For example, if an airport is caught in a blizzard, the number of arriving flights that are canceled should increase, leading to a higher cancellation percentage. Weather controls using cancellations were created using equations \ref{eq:destcanper} and \ref{eq:origincanper}. The summary statistics for each control are provided in Table~\ref{tab:sumstatscontrol}.

\begin{equation}
    \textit{Destination \%} = \left( \frac{\text{number of canceled flights}_{\text{destination}}}{\text{number of flights}_{\text{destination}}}\right) \times 100
    \label{eq:destcanper}
\end{equation}

\begin{equation}
    \textit{Origin \%} = \left( \frac{\text{number of canceled flights}_{\text{origin}}}{\text{number of flights}_{\text{origin}}}\right) \times 100
    \label{eq:origincanper}
\end{equation}

\begin{table}[h!]
\caption{Weather Controls Summary Statistics}
\label{tab:sumstatscontrol}

\centering
\begin{tabular}{L{2.5cm} L{4cm} R{2cm} R{2cm} R{2cm} R{2cm}}
\toprule
\textbf{Variable} & \textbf{Description} & \textbf{Minimum} & \textbf{Maximum} & \textbf{Mean} & \textbf{SD} \\
\midrule
\textit{destcanper} & The percent of flights canceled at a given destination on a given day (\%/dest/day) & 0 & 100 & 3.05 & 5.03 \\
\addlinespace
\textit{origincanper} & The percent of flights canceled at a given origin on a given day (\%/origin/day) & 0 & 100 & 3.05 & 5.13 \\
\bottomrule
\end{tabular}
\end{table}
\FloatBarrier

Days with 100 percent cancellations include December 11th 2000 when a large snowstorm swept the midwest. Midway airport in Chicago received 13.6 inches of snow that day \parencite{noauthor_december_2000}. Of the 145 arriving flights at Midway, 71.7\% were canceled leading to a \textit{destcanper} value of 71.7.

\subsection{Additional Data Used}
\begin{figure}[h!]
\caption{Total Airline Passengers Each Month}
\label{fig:totalpass}
\centering
\includegraphics[width=0.9\textwidth]{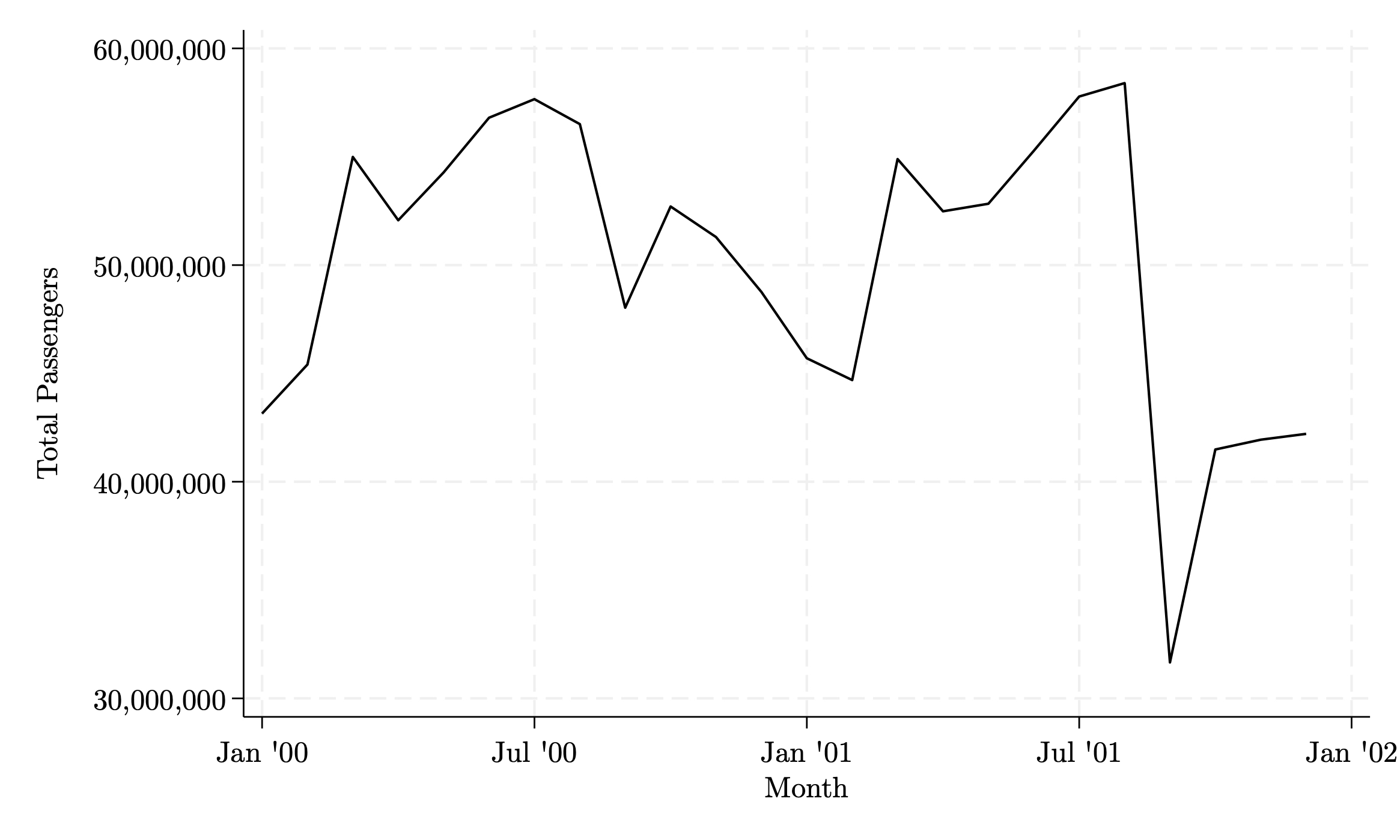}
\end{figure}

Calculating the consumer welfare gains from the decision to turn off SA requires data on passenger counts as well as the passengers' value of time. The passenger data for the years 2000 and 2001 came from BTS’s T-100 dataset for all carriers in the domestic market \parencite{noauthor_t-100_nodate}. The total number of passengers reported in the dataset were 621,657,128 in 2000 and 579,353,026 for 2001. 2001 saw less passengers due to reduced demand for flights after 9/11. My incorporation of passenger data stops short of September 2001 so this drop in demand does not affect the analysis. The breakdown by month can be seen in Figure~\ref{fig:totalpass}.

\FloatBarrier

The DOT’s economists provide a path to understand how much consumers valued the opportunity cost of the time they spent on the airplane. They publish an updated assessment every few years of the valuation consumers assign to the time they spend on various forms of transportation. One such assessment was published in 2003, however their calculations were based on 2000 numbers, the exact period this paper is examining. In the report, they assign a value of \$28.60 per hour for an average consumer traveling by air \parencite{frankel_2003_2003}. According to the report, this valuation could be as low as \$23.80 and as high as \$35.60. The interpretation of these numbers is that an average consumer is content to forgo up to \$28.60 in wages in order to spend time flying.

\section{Methodology}
In order to assign a monetary value to the effect on airline delays of turning off SA, I separate my analysis into two processes. The first uses the flight delays data to quantify the reduction in delays per flight. The second then translates the savings into consumer welfare using passengers’ valuations of their travel time. 

\subsection{Quantifying the Time Saved From Improved GPS}
To start, I investigate whether there is evidence of a relationship between the removal of GPS SA and in-flight delays in 2000. I run a simple regression looking at the effect of differentiating between airlines on additional in-flight delays compared to the effect after SA was disabled. I then plot the coefficients in order to observe whether there was a noticeable decline in each airline’s delays. The existence of a relationship motivates the rest of the analysis.

Next, I apply a more rigorous test to isolate the effect of the discrete GPS improvement on in-flight delays. I utilize a difference-in-differences econometric model, where the removal of GPS SA was the treatment applied. The group I test is all flights in 2000. Due to the lack of a second United States to use as a control, my control group is all flights in the preceding year, 1999. The regression equation is given by Equation~\ref{eq:basereg}. To delineate when the treatment, disabling GPS SA, was applied, I create a dummy variable, entitled $GPS$, which has a value of 1 between May 2nd and December 31st of each year and is otherwise equal to 0. I also create a variable, entitled $treated$, to distinguish between the treated group, flights in 2000, and my control group, flights in 1999. Finally, the cancellation controls are added to the regression and delineated by the variable $X$. The subscript $i$ indicates a given flight in the dataset while $t$ refers to the period, 1999 or 2000, in which the flight occurred.  
\begin{equation}
    adddelay_{i,t} = \beta_0 + \beta_1treated_t+\beta_2GPS_i+\beta_3(treated\cdot GPS)_{i,t}+\xi X_{i,t} +\varepsilon_{i,t}
    \label{eq:basereg}
\end{equation}

The benefit of the difference-in-differences approach is that it prevents variables that affect both 1999 and 2000 from affecting the outcome. One of the main confounding variables that this approach prevents is the seasonality of airline delays. By regressing over the full year, this approach accounts for all the seasonal rises and falls that occur in a given year.

In the analysis, $\beta_3$ is interpreted as the change in in-flight delays, in minutes, that is attributable to the removal of Selective Availability. My hypothesis predicts that this will be a negative, significant number. In other words, I expect additional in-flight delays to drop as a result of the improved accuracy of GPS.

There are a few crucial assumptions that allow this methodology to work. The first assumption is that flight delays were functionally similar in 1999 and 2000. Delays could drop between two subsequent years for a number of reasons. These include a shock to the airline industry, such as the terrorist attacks on 9/11, which led to new practices and procedures, or a large policy shift, such as the Airline Deregulation Act of 1978. In researching 1999 and 2000, it appears that there was not a significant alteration in how air travel operated that would affect the analysis. If anything, GPS becoming a usable navigation tool was the disruptive force in those two years.

The other main assumption is that all flights benefited from the improved GPS signal immediately. It is possible that some planes may not have had GPS receivers or pilots may have not had the proper training on the system. However, the history of GPS indicates that pilots did have access to the system and used it to some extent. GPS was originally opened to commercial and civilian uses in 1983 as a way of helping planes avoid flying into hostile territory \parencite{pellerin_united_2013}. Additionally, in 1994 Fiji worked with the FAA to incorporate GPS fully into its aviation system, despite the reduced civilian accuracy \parencite{stodola_how_2019}. Airlines were the originally envisioned commercial users of GPS so it is quite likely that they had started implementing it before 2000.

The initial difference-in-differences regression gives the effect over the first year averaged across all flights in the dataset. However, improved GPS probably affected delays differently depending on characteristics such as the recency to the improvement and the length of the flight. Splitting the data up into these categories will give more insight into where the effect of removing SA was most salient.

The first division I make is by the recency of the flight to the treatment. While regressing over the full years of 1999 and 2000 has a few benefits, namely avoiding seasonality and giving the effect of removing GPS SA time to show up in the data, it obscures insights into how quickly it took pilots to start improving their flight times. To investigate the effect over time, I run 4 additional regressions. The periods of the regressions are given in Table~\ref{tab:recencygroups}. They are named based on how many months are included in the group post treatment. The control periods for the regressions are the dates in 1999 and the treatment periods are the dates in 2000.

\begin{table}[h!]
    \centering
    \caption{Periods of Recency Based Regressions}
    \label{tab:recencygroups}
    \begin{tabular}{llll}
    \toprule
    \textbf{Name} & \textbf{Starting Date} & \textbf{Treatment Date} & \textbf{Ending Date} \\
    \midrule
    1 month & April 1st & May 2nd & May 31st \\
    2 months & March 1st & May 2nd & June 30th \\
    3 months & February 1st & May 2nd & July 31st \\
    4 months & January 1st & May 2nd & August 31st \\
    \bottomrule
    \end{tabular}
\end{table}

I expect to see the treatment effect either grow or shrink as the data incorporates flights further from the treatment. A possible reason that delay savings would increase over time is due to pilots learning to use the now accurate GPS. As pilots spent more time with the GPS, they may have become more comfortable relying on it. In contrast, the effect may decline as airlines and the FAA adapted to it and created parameters on how the more precise navigation could be used. 

Another way that the effect may vary is based on the distance of the flights. Unfortunately, the dataset does not provide the exact distance of each flight so I cannot investigate if the improved accuracy of GPS led to straighter flight paths; however, having the distance between the airports allows me to test if the effect of removing SA varied with the length of the flights. I expect that the effects of improved GPS will be more salient as flight distances increase since there is more flight path to optimize. This finding would help validate that the mechanism through which GPS reduces delays is by aiding pilots in adhering to their flight paths. In order to test this, I divide the data into 6 groups of 500 miles each.\footnote{The groups are 0-500, 500-1000, 1000-1500, 1500-2000, 2000-2500, 2500+}

It is possible that the delay savings across distance groups changed at different rates. Typically, longer routes tend to be piloted by more experienced pilots. These pilots may have adapted to the accurate GPS faster leading to increased savings over time. To investigate this, I rerun the regression for each distance group within each of the four month subsets.

In order to investigate one other possible area in which improving GPS had an effect, I run a separate difference-in-differences regression, given by Equation~\ref{eq:distancereg}, where $distance$ is the dependent variable. It is possible that airlines, responding to increased efficiency from flying with GPS, changed the routes they were flying. This would cause a significant change in the flight distances.
\begin{equation}
    distance_{i,t} = \beta_0 + \beta_1treated_t+\beta_2GPS_i+\beta_3(treated\cdot GPS)_{i,t}+\xi X_{i,t} +\varepsilon_{i,t}
    \label{eq:distancereg}
\end{equation}

I expect that the results of this regression will not be significant due to the time it takes to implement new routes. If airlines responded to the efficiency improvements in their routing decisions, it was probably not immediately, and thus would not show up in the data.

\subsection{Quantifying the Monetary Value of Time Saved}
Consumer welfare gains from the reduced delays can be found by summing how much each passenger valued the time they saved in-flight. For each of the four types of regressions -- initial, by recency, by distance, and by both recency and distance -- I use the T-100 dataset to find the number of passengers that were in the respective groups. While the improvements may only save each passenger a small amount of time, cumulatively all passengers experience a large welfare increase. Therefore multiplying those passenger numbers by the $\beta_3$ coefficients from each regression yields how much time passengers saved from the reduced delays. To convert this into a monetary amount that can be compared to other policy actions, I multiply the amount of time saved by the DOT’s estimates for how much airline passengers value their time. I also calculate the savings based on the DOT’s upper and lower bounds on these value of time estimates.

\section{Results}
As outlined in my methodology, I started by running regressions to find the effect of the GPS improvement treatment in minutes per flight. I tested the treatment on subgroups of the data in order to map how the treatment evolved over time and how it affected longer flights. I then used passenger data to estimate how much consumer welfare was gained under each regression outcome. 

\subsection{GPS Effects on In-flight Delays}
\begin{figure}[h!]
\caption{Coefficients Before and After SA Removal}
\label{fig:coefplot}
\centering
\includegraphics[width=0.9\textwidth]{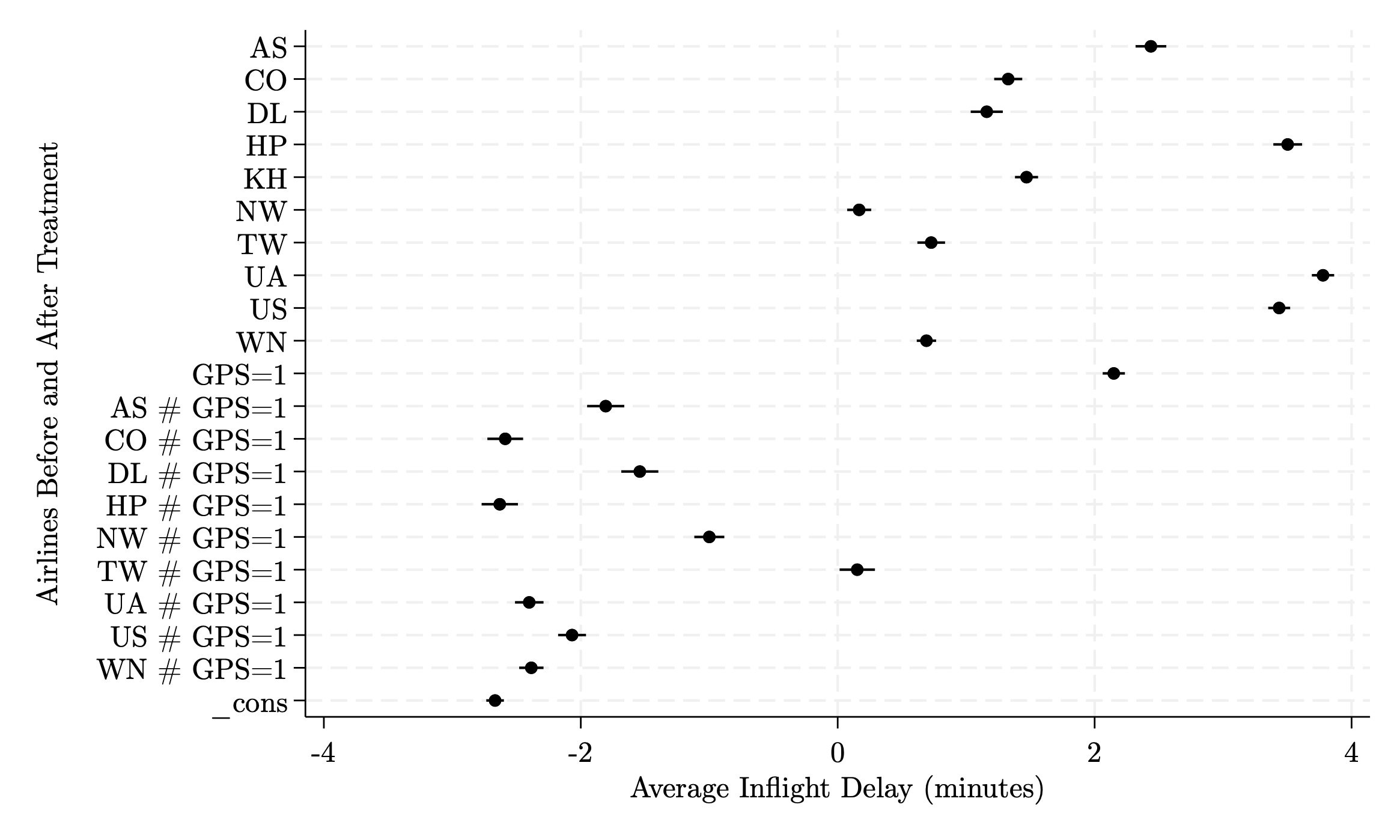}
\end{figure}

The first step was to test if there is an observable relationship between the implementation of GPS and additional in-flight delay. I ran a regression over the 2000 data stratified by airline and observed each airline's coefficients before and after the removal of SA. Plotting the coefficients in Figure~\ref{fig:coefplot} shows that after the GPS signal improved, the additional delays gained in-flight dropped and even became negative in most cases. While this relationship is not rigorous, it indicates that the change in GPS policy is correlated with a decrease in delays and is worth investigating.

\FloatBarrier

The next step was running the difference-in-differences regression with the proxy weather controls. The control group contained all flights in 1999, while the treated group was made up of all flights in 2000. The results are given in Table~\ref{tab:initialreg}.

\begin{table}[h!]
    \centering
    \caption{Initial Regression on In-flight Delays}
    \label{tab:initialreg}
\begin{tabular}{lc} \hline
 & (1) \\
VARIABLES & adddelay\\ \hline
 &  \\
treated & 0.342***\\
 & (0.0142)\\
 \addlinespace
GPS & 0.962***\\
 & (0.0126) \\
 \addlinespace
treated\#GPS & -0.617***\\
 & (0.0175)\\
 \addlinespace
destcanper & 0.374***\\
 & (0.00213)\\
 \addlinespace
origincanper & 0.295***\\
 & (0.00227)\\
 \addlinespace
Constant & -3.271***\\
 & (0.0117)\\
 &  \\
Observations & 10,840,690\\
 R-squared & 0.024\\ \hline
\multicolumn{2}{c}{ Robust standard errors in parentheses} \\
\multicolumn{2}{c}{ *** p$<$0.01, ** p$<$0.05, * p$<$0.1} \\
\end{tabular}
\end{table}
\FloatBarrier

The important coefficient in this regression is on the interaction between $treated$ and $GPS$. This coefficient equals -0.617 and has a $t$ value of -35.17, i.e. it is very significant.\footnote{ The regressions throughout this section have low $R^2$ values, mostly around 0.02-0.03. This is to be expected from this type of analysis. There are many reasons that flights are delayed, which means that the treatment, removal of SA, only explains a small fraction of overall delays. This analysis is intended to only quantify the change in delays from removing SA, not predict or explain all delays. Therefore, in this case, significant findings are more important than a high $R^2$.} 
The interpretation of -0.617 is that the removal of GPS SA from the treated group, flights in 2000, led to a decrease of 0.617 minutes in average in-flight delays per flight. While this number appears insignificantly small, once scaled up by number of flights or number of passengers, it grows considerably. In the rest of 2000, after May 2nd, there were 3,805,944 flights. Each one saved, on average, 0.617 minutes which adds up to 39,137 hours of flight time saved in just 8 months. 

It is important to investigate whether 0.617 minutes per flight can be further optimized since each flight was not identical. Increased optimization will lead to a more accurate portrayal of the cumulative effect of GPS as a global public good. Therefore, to investigate whether a more targeted analysis could yield further insights, I ran additional regressions as outlined in my methodology. I started by changing the periods over which I ran the regressions to capture whether the time elapsed since the treatment influenced its effect. The regression results are in Table~\ref{tab:recencyregs}.

\begin{table}[h!]
    \centering
    \caption{Regressions with Varied Periods}
    \label{tab:recencyregs}
\begin{tabular}{lcccc} \hline
 & 1 Month & 2 Months & 3 Months & 4 Months \\
VARIABLES & adddelay & adddelay & adddelay & adddelay \\ \hline
 &  &  &  &  \\
treated & -0.997*** & 0.0344* & 0.458*** & 0.351*** \\
 & (0.0285) & (0.0194) & (0.0158) & (0.0142) \\
\addlinespace
GPS & -0.358*** & 1.264*** & 1.738*** & 1.740*** \\
 & (0.0303) & (0.0213) & (0.0175) & (0.0152) \\
\addlinespace
treated\#GPS & -0.434*** & -1.508*** & -1.687*** & -1.185*** \\
 & (0.0430) & (0.0304) & (0.0245) & (0.0211) \\
\addlinespace
destcanper & 0.519*** & 0.473*** & 0.456*** & 0.400*** \\
 & (0.00628) & (0.00410) & (0.00308) & (0.00263) \\
\addlinespace
origincanper & 0.463*** & 0.434*** & 0.351*** & 0.320*** \\
 & (0.00785) & (0.00477) & (0.00344) & (0.00272) \\
\addlinespace
Constant & -2.185*** & -3.062*** & -3.448*** & -3.423*** \\
 & (0.0239) & (0.0165) & (0.0138) & (0.0124) \\
 &  &  &  &  \\
Observations & 1,801,160 & 3,622,628 & 5,377,074 & 7,188,590 \\
R-squared & 0.036 & 0.033 & 0.030 & 0.029 \\ \hline
\multicolumn{5}{c}{Robust standard errors in parentheses} \\
\multicolumn{5}{c}{*** p$<$0.01, ** p$<$0.05, * p$<$0.1} \\
\end{tabular}
\end{table}
\FloatBarrier

The coefficients of interest for the 1, 2, 3, and 4 month regressions respectively are -0.434, -1.508, -1.687, and -1.185 minutes per flight. They were all very significant with p values less than 0.00. A potential reason for the increasing values is that pilots started to adapt to the improved GPS signal and thus took better advantage of it as time went on. However, the improvement starts to decline after the 3 month regression and by the time of the year long regression the effect had decreased to -0.617 minutes per flight. This decline appears to invalidate the theory that delays would continue to drop as pilots spent more time with the improved GPS system. An alternate interpretation, that builds upon pilots getting more familiar with the system, is that pilots used their newfound mastery of the system to help airlines and regulators put guidelines into place that may have made its use less efficient but more safe. This would explain the gradual decline of GPS’s effect on delays. 

Another way that I refined the analysis was by splitting the flights up by their length in miles. It is possible that the improved GPS had more of an effect over longer distances because it reduced delays by mile and longer flights cover more miles. The results of the regressions by flight distance are listed in Table~\ref{tab:distanceregs}. 

\begin{table}[h!]
    \centering
    \caption{Regressions with Data Subdivided by Flight Distance}
    \label{tab:distanceregs}
\begin{tabular}{lcccccc} \hline
 & 0--500 & 500--1000 & 1000--1500 & 1500--2000 & 2000--2500 & 2500+ \\
VARIABLES & adddelay & adddelay & adddelay & adddelay & adddelay & adddelay \\ \hline
 &  &  &  &  &  &  \\
treated & 0.220*** & 0.290*** & 0.573*** & 0.153** & 1.750*** & 2.581*** \\
 & (0.0174) & (0.0251) & (0.0450) & (0.0695) & (0.107) & (0.200) \\
\addlinespace
GPS & 0.591*** & 1.088*** & 1.596*** & 0.872*** & 2.524*** & 2.787*** \\
 & (0.0154) & (0.0227) & (0.0402) & (0.0610) & (0.0947) & (0.170) \\
\addlinespace
treated\#GPS & -0.554*** & -0.496*** & -0.699*** & -0.457*** & -2.883*** & -2.408*** \\
 & (0.0216) & (0.0315) & (0.0560) & (0.0839) & (0.131) & (0.238) \\
\addlinespace
destcanper & 0.294*** & 0.466*** & 0.417*** & 0.403*** & 0.432*** & 0.710*** \\
 & (0.00268) & (0.00402) & (0.00642) & (0.00939) & (0.0117) & (0.0192) \\
\addlinespace
origincanper & 0.245*** & 0.321*** & 0.368*** & 0.410*** & 0.470*** & 0.192*** \\
 & (0.00283) & (0.00416) & (0.00689) & (0.00978) & (0.0154) & (0.0201) \\
\addlinespace
Constant & -1.860*** & -3.710*** & -4.982*** & -5.186*** & -6.812*** & -8.697*** \\
 & (0.0141) & (0.0213) & (0.0380) & (0.0604) & (0.0956) & (0.172) \\
 &  &  &  &  &  &  \\
Observations & 4,575,387 & 3,569,307 & 1,448,332 & 731,150 & 376,778 & 139,736 \\
R-squared & 0.025 & 0.032 & 0.022 & 0.021 & 0.017 & 0.026 \\ \hline
\multicolumn{7}{c}{Robust standard errors in parentheses} \\
\multicolumn{7}{c}{*** p$<$0.01, ** p$<$0.05, * p$<$0.1} \\
\end{tabular}
\end{table}
\FloatBarrier

In these regressions, the amount of time saved per flight as a result of GPS is around 0.5 for all flights less than 2000 miles in distance. After 2000 miles, the time saved per flight jumps to 2.883 minutes. These results seem to validate the proposition that GPS saves time per mile flown, resulting in more savings on longer flights. That being said, this explanation would also predict a linear rise in the coefficients as distance increases, which is not evident in the regressions. The lack of a linear relationship may indicate that the grouping bins of 500 miles are not the best to explore this relationship or that there is an alternate explanation for these results. 

\begin{figure}[h!]
\caption{Average Regression Coefficients for Flights Over and Under 2000 Miles}
\label{fig:over/under2000}
\centering
\includegraphics[width=1\textwidth]{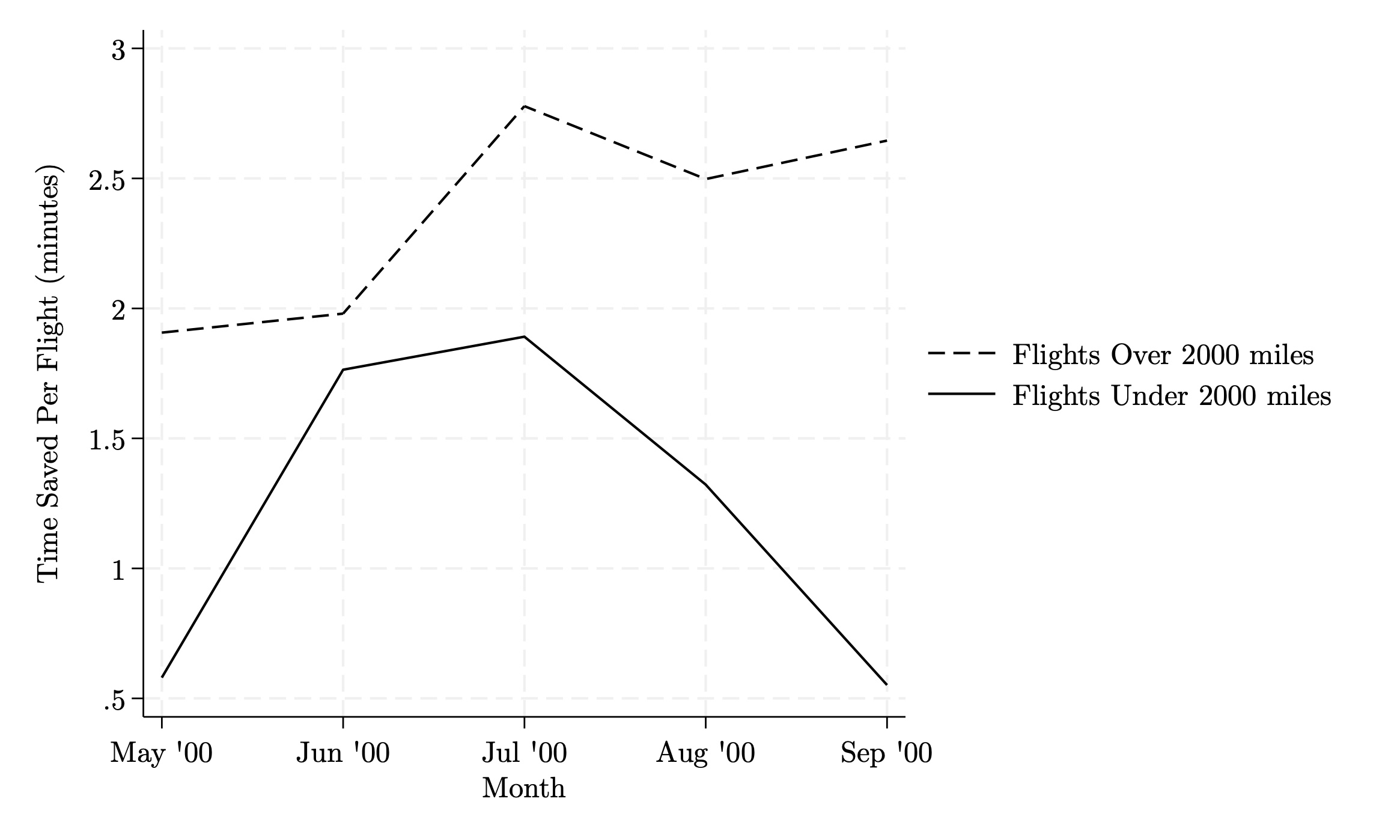}
\end{figure}

To find the most accurate treatment effects, I combined the recency regressions and distance regressions. To achieve this, I reran the distance based regressions within the 1, 2, 3, and 4 month subdivisions that I used to test the recency effects. The results of these regressions are given in full in Appendix~\ref{sec:regdistrecency}. All but one of the coefficients were statistically significant with most at the 1\% level. The treatment effects from the regressions are shown in Figure~\ref{fig:over/under2000}. They are aggregated into flights under 2000 miles and flights over 2000 miles because these two groupings exhibited similar patterns within the groups. 

\FloatBarrier

Figure~\ref{fig:over/under2000} shows how the treatment effects diverged based on whether the flight was short or long range. The longer flights, above 2000 miles, rose to values above 2 minutes saved per flight and stayed there over the 5 months. This is potentially a factor of pilot skill. More senior pilots often choose longer flights so perhaps these pilots were able to, or more willing to, take advantage of the improved GPS signals. For the rest of the flights, between 0 and 2000 miles in length, the minutes saved initially rise and then fall down to around a half of a minute. One possible explanation for this is that as pilots learned how to use the system, they created guidelines which reduced efficiency but increased safety. Another explanation is that the months in question are during the summer, a period of high delays. Thus the efficiency gains were higher relative to the size of the typical delays. This explanation however, does not hold for the longer distance flights. 

To not overlook a path through which disabling selective availability changed airline behavior, I regressed flight distances on my variables. The $distance$ variable in the dataset records the miles between the origin and destination airports, not the actual miles flown by the plane. The variable values are thus controlled by the airlines, who set the routes, and not the pilots, who fly them. It is possible that these airlines changed the distances or distribution of routes in order to take advantage of the ability to follow flight plans with greater accuracy. The full regression results are given in Table~\ref{tab:distancevarreg}.

\begin{table}[h!]
    \centering
    \caption{Regression with $distance$ as the Dependent Variable}
    \label{tab:distancevarreg}
\begin{tabular}{lc} \hline
 & (1) \\
VARIABLES & distance \\ \hline
 &  \\
treated & 9.125*** \\
 & (0.584) \\
\addlinespace
GPS & 11.94*** \\
 & (0.509) \\
\addlinespace
treated\#GPS & 0.902 \\
 & (0.719) \\
\addlinespace
destcanper & 1.663*** \\
 & (0.0379) \\
\addlinespace
origincanper & 1.170*** \\
 & (0.0369) \\
\addlinespace
Constant & 737.3*** \\
 & (0.434) \\
 &  \\
Observations & 11,210,300 \\
R-squared & 0.001 \\ \hline
\multicolumn{2}{c}{Robust standard errors in parentheses} \\
\multicolumn{2}{c}{*** p$<$0.01, ** p$<$0.05, * p$<$0.1} \\
\end{tabular}
\end{table}
\FloatBarrier

As expected, the coefficient on the interacted term is not significant. This means that airlines likely did not alter flight routes as a result of GPS improving, at least not in the first year or as reflected in the available data. 

I tested a couple of other alterations to the regressions. One way I altered the regressions was by changing the periods such that the treated period was from November 1999 to the end of October 2000 and the control period was from November 1998 to the end of October 1999. This had the effect of creating an equal number of months before and after the treatment was applied. In the initial regression setup, these new periods yielded a coefficient of -0.640. This value falls within the 95\% confidence interval of the original regression output, -0.617. These outcomes were close enough that I ultimately decided not to use this design for the rest of the regressions or results. I also tested other controls such as airline, plane model, and season. I did not include these controls in my final analysis because they either were missing data or did not meaningfully impact the outcomes.

\subsection{Estimated Welfare Effects}
Using the most segmented, and thus specific, estimates of the minutes of delay saved, I found that removing GPS SA led to a welfare gain of \$268 million (2000 USD) in the first year. As the regressions became more specific, by incorporating recency and flight distance, the valuations of consumer welfare gains grew in size. 

The initial regression, using the full dataset, yielded a welfare gain of \$183,464,122 over 12 months, about \$80 million less than the final amount I found. In order to find this number, I multiplied the number of passengers each month by 0.617 minutes saved per passenger and \$0.477, the DOT’s estimate of how much each passenger values a minute of air travel. Table~\ref{tab:inputsinitial} lists the passenger counts and welfare gains by month. The monetary amount for each month is equal to how much consumer welfare would have been forfeited under the less accurate GPS regime.

\begin{table}[h!]
    \centering
    \caption{Inputs to Calculating Total Consumer Valuation of Time Saved}
    \label{tab:inputsinitial}
    \begin{tabular}{L{3cm}R{4cm}R{4cm}R{4cm}}
    \toprule
    \textbf{Month} & \textbf{Domestic Passengers} & \textbf{Aggregate Time Saved (Hours)} & \textbf{Valuation of Time Saved (2000 USD)} \\
    \midrule
May ‘00 & 54,277,048 & 558,149 & \$15,963,061 \\
June ‘00 & 56,802,608 & 584,120 & \$16,705,836 \\
July ‘00 & 57,658,680 & 592,923 & \$16,957,610 \\
August ‘00 & 56,507,432 & 581,085 & \$16,619,024 \\
September ‘00 & 48,034,976 & 493,960 & \$14,127,247 \\
October ‘00 & 52,703,608 & 541,969 & \$15,500,307 \\
November ‘00 & 51,295,668 & 527,490 & \$15,086,227 \\
December ‘00 & 48,759,732 & 501,413 & \$14,340,400 \\
January ‘01 & 45,701,696 & 469,966 & \$13,441,021 \\
February ‘01 & 44,692,472 & 459,588 & \$13,144,205 \\
March ‘01 & 54,891,328 & 564,466 & \$16,143,723 \\
April ‘01 & 52,483,124 & 539,701 & \$15,435,462 \\
    \midrule
Total & 623,808,372 & 6,414,829 & \$183,464,122\\
    \bottomrule
    \end{tabular}
\end{table}
\FloatBarrier

As the table illustrates, President Clinton’s decision to end selectively availability generated at least \$183 million in additional US consumer welfare in the first year. This number is based on the DOT’s primary valuation estimate. According to the DOT, the valuation air passengers place on their time in flight could be as low as \$23.80, reducing the savings to \$152,672,940, or as high as \$35.60, increasing them to \$228,367,928. This is a \$76 million difference, equivalent to \$130 million in 2024 dollars or the entire budget request of the state of Washington’s Department of Fish and Wildlife for the 2025-2027 budget cycle \parencite{noauthor_2025_nodate, noauthor_cpi_nodate}. This kind of large scale difference stemming from adding or subtracting \$11.80 highlights the size of benefits that improvements to global public goods can bring.

The next two groups of regressions I ran were the recency and distance subdivisions. To calculate the welfare from each type of regression, I split the T-100 passenger counts either by month or by distance. For the recency regressions, I replaced 0.617 with the 1, 2, 3, and 4 month coefficients for the months of May ‘00, June ‘00, July ‘00, and August ‘00 respectively. This resulted in the calculated welfare gain rising to \$247,561,218, with a lower bound of \$206,012,482 and an upper bound of \$308,153,124. For the regressions by distance groups, I replaced 0.617 with the minutes per flight saved at each distance level. These ranged from 0.496 to 2.883. When I multiplied these values by the number of passengers that flew within each distance group and their valuation of the time, I found that removing SA led to a welfare gain of \$203,749,949 in the first year. This welfare gain is less than the one calculated from the recency regressions which indicates that fluctuations in the effect of removing SA over time had more of an effect on the welfare saved than varying the distance of the flights.  

\begin{figure}[h!]
\caption{Calculated Welfare Gains Per Month}
\label{fig:monthlywelfare}
\centering
\includegraphics[width=1\textwidth]{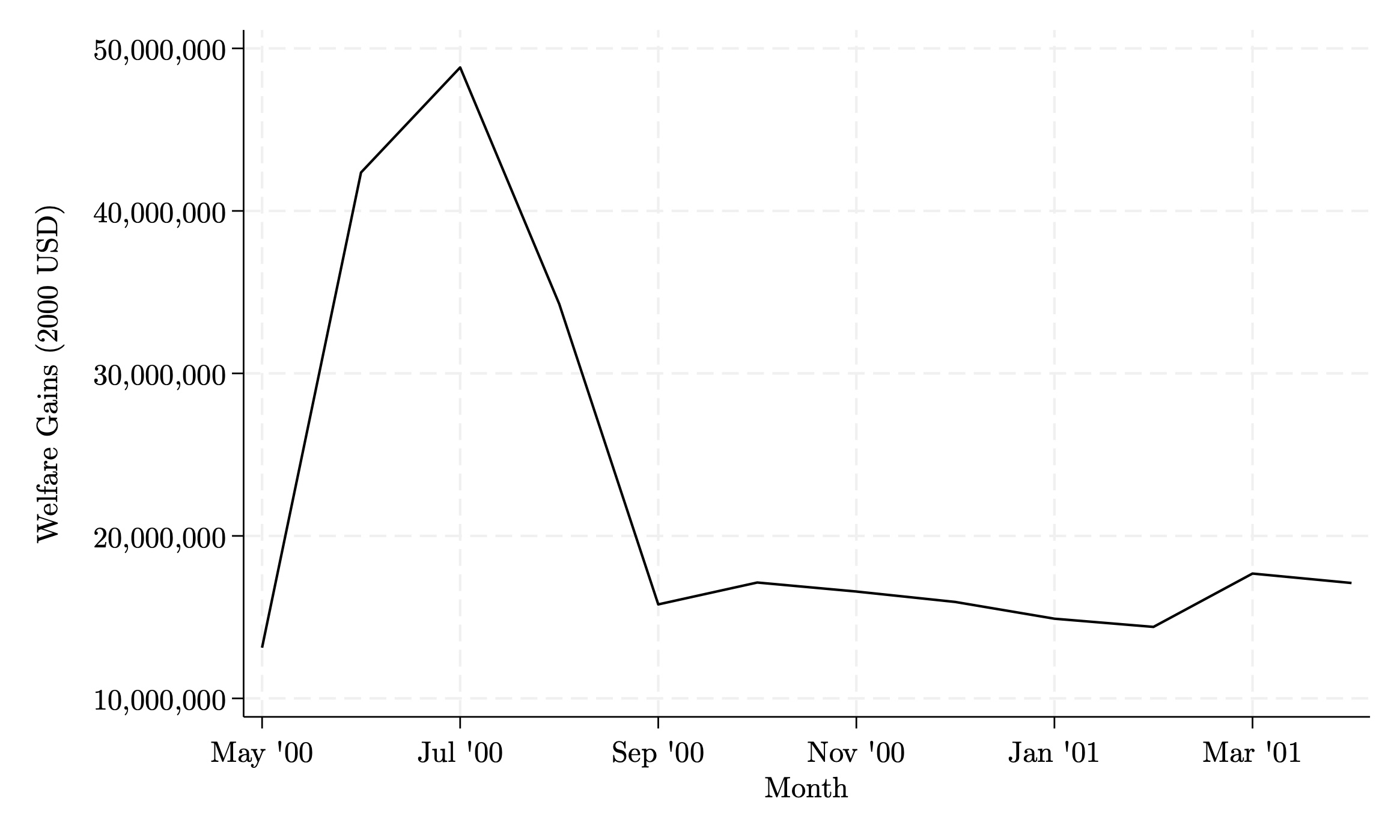}
\end{figure}

The final, and most specific, set of regressions I ran were over the distance groups within each month subdivision. These regressions yielded 6 different coefficients for each of the first 4 months. In updating the welfare gain estimate, I subdivided passengers by distance group and month and applied these coefficients for those first four months.\footnote{One of the coefficients was not significant so I replaced it with the corresponding coefficient from the distance based regressions.}
For the following 8 months, I used the treatment effects I found in the distance based regressions. The new welfare benefit calculation yielded a value of \$268,123,934 or \$223,124,113 to \$333,748,673 using the DOT’s lower bound and upper bound valuations, a difference of about \$205 million in 2024 dollars \parencite{noauthor_cpi_nodate}. The calculated welfare gains by month are plotted in Figure~\ref{fig:monthlywelfare}.

\FloatBarrier

The increased time saved in the early months combined with the increased efficiency of the model from subdividing flights by distance leads to the early peak seen in the graph. Two possible explanations for this peak are that pilots and regulators got used to having access to better GPS or that these were summer months which increased the relative impacts of more accurate GPS. Either explanation plausibly accounts for the bump in welfare in these months. For the rest of the months in 2000 and early 2001, welfare gains per month were constant around \$15 million. 

In the initial regression, the welfare benefits gained by passengers as a result of removing Selective Availability totaled around \$183 million or, at its lowest possible value, \$153 million. By running regressions over tighter periods of time, I was able to refine the results and raise the savings to just over \$247 million. Finally, I theorized that the improved GPS signal would be more salient on longer distance flights and therefore I divided the flights by distance and ran separate regressions. This work was able to raise savings to \$268 million with an upper bound of almost \$334 million.\footnote{The estimated welfare gains, including upper and lower bound, under each regression scenario are located in Appendix~\ref{sec:welfaretable}.} For a relatively costless alteration to GPS, this is a massive efficiency gain.

The benefits to US airline passengers alone paid for the GPS system for over 2 years meaning that any other benefits in other industries or around the globe were pure welfare increases. One could argue that turning off SA had a negative national security externality by allowing adversaries to access more accurate positioning, and thus targeting, signals. However, \textcite{weeden_is_2013} finds in his cost benefit analysis that GPS was already sufficient enough for most weapons targeting and thus the removal of SA did not result in a large negative national security externality.

\subsection{Challenges to GPS Provision: Jamming and Spoofing}
In the same way that GPS improvements led to large welfare benefits, interruptions to GPS can quickly wipe out commensurate amounts of societal welfare. Two prominent ways that GPS can be easily interrupted is through jamming and spoofing. Jamming is when an actor electronically blocks out GPS's signals, denying access to them for anyone in the area. Spoofing is the falsification of GPS signals to misrepresent the position of a GPS receiver. This can lead to planes or missiles flying off course without realizing their error. Both spoofing and jamming used to be relatively rare, however they have become increasingly common since 2023 \parencite{opsgroup_gps_2024}. According to a report by the OPSGROUP, spoofing incidents increased by 500\% in 2024. This increase was mainly due to the proliferation of conflict and modern drone warfare, especially in the regions around Israel, Ukraine, and the India/Pakistan border \parencite[p.~14]{opsgroup_gps_2024}. 

While civilian airliners are not the main target of these interferences, their GPS receivers are affected nonetheless. The OPSGROUP found that 1,500 flights were spoofed a day in 2024 with 41,000 affected in just one month from July to August. The welfare loss from these spoofing incidents can be calculated using a method similar to what I used in my analysis of SA. According to BTS, there were, on average, 107 passengers per flight in those two months \parencite{noauthor_data_nodate}. DOT’s most recent update to their passenger value of time estimates was in 2014 and equaled \$60.64 per hour, in 2024 dollars \parencite{rogoff_revised_2014,noauthor_cpi_nodate}. Estimating the increased delay from spoofing is harder because pilots do not always know if they are being spoofed and may drift off course for an extended period of time before correcting. In Table~\ref{tab:spoofingwelfareloss}, I calculate the passenger welfare lost in the one month period identified by OPSGROUP at different levels of increased delay. I assume that each of the 41,000 affected flights held 107 passengers who valued their travel time at \$60.64 per hour.

\begin{table}[h!]
    \centering
    \caption{Welfare Losses at Increasing Levels of In-Flight Delay}
    \label{tab:spoofingwelfareloss}
    \begin{tabular}{R{3.5cm}R{3.5cm}}
    \toprule
    \textbf{Additional Delay (Minutes)} & \textbf{Welfare Lost in One Month} \\
    \midrule
    0.617 & \$2,735,651\\
    \addlinespace
    5&\$22,168,973 \\
    \addlinespace
    25&\$110,844,867 \\
    \addlinespace
    60&\$266,027,680 \\
    \addlinespace
    120&\$532,055,360 \\
    \bottomrule
    \end{tabular}
\end{table}
\FloatBarrier

As Table~\ref{tab:spoofingwelfareloss} shows, spoofing likely caused millions of dollars of welfare loss, even at very small levels of per-flight delay. It is very probable, however, that the delay for each flight was more substantial than the delay saved in 2000 when SA was removed. In 2000, pilots had the experience of flying without GPS and were just learning how to fly with it. Nowadays, pilots often fly using GPS so they are probably less accustomed to flying without it. Thus, when a GPS signal is spoofed or jammed, pilots have less experience adapting, resulting in higher in-flight delays. More rigorous quantitative work is needed to know the true scale of these disruptions in terms of flight time and welfare losses.

\section{Policy Recommendations and Future Research}

President Clinton’s press release on May 1st, 2000 promised that the removal of SA would bring “immediate” and “tangible” benefits to users “around the world” \parencite{noauthor_president_2000}. This analysis shows that President Clinton did not lie. US airline passengers reaped \$13 million in welfare benefits in the first month after this statement and a total of \$268 million in the ensuing year. However, President Clinton’s vision for GPS was as a global public good, benefiting people in every country. Because of the vast and diffuse nature of global public goods, like GPS, any efficiency improvement scales with population meaning that the welfare gains from removing SA were probably in the billions on a global level. Multiple countries, such as Fiji, were already using GPS in their national aviation systems and would have instantly benefited along with the US. 

As the removal of Selective Availability highlights, GPS contributes to massive global welfare. Below, I discuss ways to combat the increasing proliferation of jamming and spoofing threats. Additionally, I propose avenues to build upon this research and further refine our understanding of the economics benefits gained with GPS.

\subsection{Policy Recommendations: Jamming and Spoofing}

The costs of disrupting GPS signals extend beyond airline navigation. Many industries now rely on GPS to varying extents. Even industries that do not require navigation, like banking and stock trading, use GPS for its accurate timing signals. \textcite{oconnor_economic_2019} place the cost of disruption at \$1 billion a day, in the US, while acknowledging that this number is an underestimate. Now that spoofing and jamming are becoming more commonplace, countries near conflict zones are feeling this economic loss on a regular basis. The loss or diversion of GPS signals also can lead to loss of life. Policy makers need to seriously address the threats to GPS posed by jamming and spoofing to avoid large potential human and economic costs.

Multiple organizations are tracking the human costs of modern GPS disruptions. In their report, OPSGROUP concluded that “the impact of GPS Spoofing on flight safety… is extremely significant” (p.~5). Just recently, GPS jamming by the Russian military prevented an Azerbaijan Airlines flight from landing at its intended destination. On its return to its origin airport, the flight was struck by a missile and crashed \parencite{petchenik_azerbaijan_2024}. While jamming was not the direct cause of the crash, it contributed to the conditions under which the airplane ended up in an active conflict zone. In addition, the International Committee of the Red Cross (ICRC) has identified disruptions to space-based systems, like GPS, as impediments to humanitarian work and emergency response \parencite{noauthor_preliminary_2023}. In a working paper, ICRC presented 5 recommendations to mitigate the effect of space warfare. One recommendation is that states should not engage in any military acts that disrupt “space systems necessary for the provision of essential civilian services,” i.e. space-based GPGs \parencite[p.~3]{noauthor_preliminary_2023}. They describe the consequences of these disruptions to space-based GPGs as “far-reaching,” similar in scale to the large and diffuse impact from the removal of SA \parencite[p.~3]{noauthor_preliminary_2023}.

One solution to the adverse costs from GPS disruptions is to, as the ICRC recommends, ban the use of jamming and spoofing in conflicts. While this would solve the current problem, it seems unlikely that states will agree to withhold their use of these powerful tools. In the conflict zones where GPS jamming and spoofing are most heavily used, the states that employ them, such as Israel, Russia, and Ukraine, are in close proximity to their targets and are therefore also impacted by the negative welfare externalities. The fact that these states continue to deploy GPS jamming and spoofing technology is evidence that they find it militarily worthwhile and thus would probably not agree to a blanket ban. 

Therefore, another way to mitigate the effect of GPS disruptions is to invest in improvements to the system. As the decision to remove SA showed, government investment into GPS yields very large scale welfare benefits. One current proposal is to field “Resilient GPS” (R-GPS). R-GPS is intended to be a constellation of many small satellites designed to supplement the current GPS signal \parencite{noauthor_space_2024}. The program is expected to eventually cost around \$1.2 billion, although military acquisition has tended to end up over budget in the last few decades \parencite{erwin_sierra_2025}. This price tag, while appearing high, is equal to the welfare lost in just 2 days of a complete GPS blackout, according to the calculations done by \textcite{oconnor_economic_2019}. If the R-GPS program succeeds, it will enable to US to continue providing unobstructed access to GPS signals, preserving the welfare generated by the system. However the program has a long way to go with some already pointing out flaws in the acquisition goals \parencite{mcgivern_resilient_2024}.

Regardless of the solution chosen, the work in this paper demonstrates that it is worthwhile for policy makers to invest in ways to counter GPS disruptions. The global welfare benefits from mitigating these disruptions are, at minimum, in the hundreds of millions of dollars annually. Additionally, the military should not monopolize access to these GPS improvements. GPGs are able to provide such large welfare gains because of the sheer number of people who benefit. If the military employs a similar policy to Selective Availability, it will deny the global economy these large and potentially life-saving welfare gains.

\subsection{Limitations and Future Research}
There are a few controls that were not included for feasibility reasons that would be interesting to study as a continuation of this work. For one, acquiring the actual weather data from NOAA would shed light on whether the proxy weather method was an adequate approach to controlling for weather. Integrating this data is technically difficult because of the sheer volume required, but its addition could lead to richer results. It could also be interesting to add controls for spending behavior. I did not consider whether changes in consumer spending habits affected traffic at airports. It is possible consumers started spending less in 2000 and therefore there was less congestion in the airspace. Additionally, recent journalism has focused on how airlines change their scheduled times to reduce the reported delays, even if planes fly slower \parencite{blatt_airlines_2024}. Due to the short time frame and methodology of the analysis, this was most likely not a factor in the results but it may be worth investigating. 

\begin{figure}[h!]
\caption{Average Quarterly Ticket Fares 1999 - 2000}
\label{fig:DB1Bfares}
\centering
\includegraphics[width=1\textwidth]{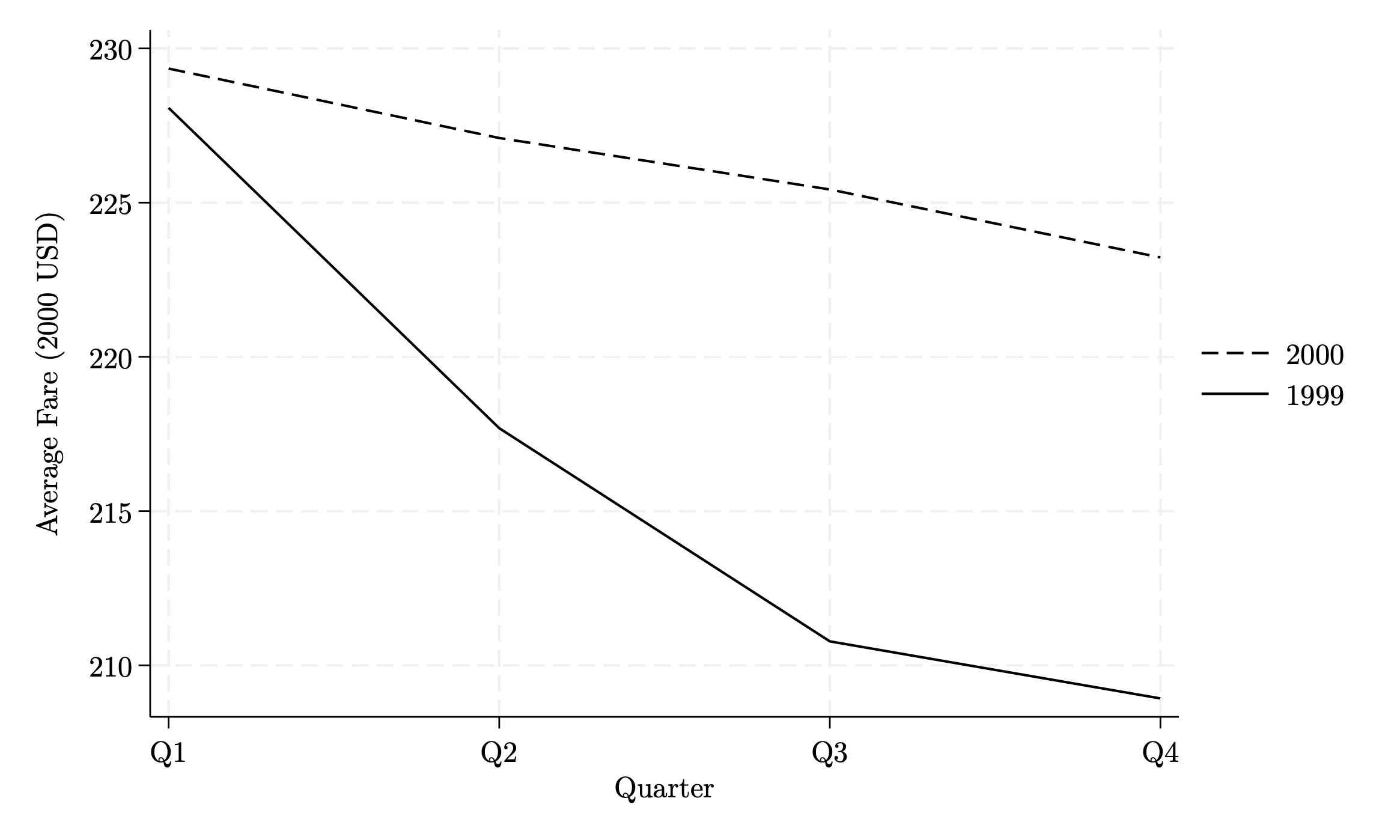}
\end{figure}

Another important area of research that builds on this paper is the distribution of welfare gains. I assume throughout the paper that consumers captured almost all of the rents from the reduced delays. However, airlines may have raised their prices after the removal of SA in order to capture this rent for themselves. In this case, the political economy of GPS improvements would change. If firms are primarily capturing the rents, they may pressure the government to protect GPS in a more coordinated and powerful fashion than consumers could. A preliminary analysis of the BTS’s DB1B database, a 10\% sample of ticket prices each year, shows a potential relationship between ticket fares and the removal of SA. Figure~\ref{fig:DB1Bfares} depicts average ticket fare per quarter in the dataset.

\FloatBarrier

The treatment, removing Selective Availability, occurred midway through the 2nd quarter (Q2) of 2000. The dataset does not distinguish flights by date so it is impossible to determine which Q2 flights were before the treatment and which were after. However, Figure~\ref{fig:DB1Bfares} appears to show that the roughly \$20 decline in ticket fares that happened over 1999 did not repeat itself in 2000. Instead the average fare dropped just over \$6 between quarters 1 and 4 of 2000. While this is not a rigorous analysis, it does indicate a potential relationship. If ticket prices did rise as a result of removing SA, then producers captured some of the welfare generated by the policy action. Future research on this topic would have implications for how governments understand the domestic dynamics of GPG improvements. 

This work could also be expanded to examine the global welfare gains from the removal of SA. While SA was mainly disabled for domestic reasons, it did have considerable spillover effects to the rest of the world. Additionally, President Clinton indicated that the global impact was part of his decision making \parencite{noauthor_president_2000}. Knowing the share of benefits felt in the US versus in the rest of the world would shed light on the work done in \textcite{boadway_country_1999} on domestic rationales for GPG provision. They posit that governments’ willingness to contribute to public goods rises as the domestic benefit increases. If a large percentage of GPS’s total benefits are felt in the US, policy makers may be more likely to support investing in the system because it has a disproportionate domestic benefit. The main difficulty in conducting this global analysis is data availability. There is no centralized global reporting system for airline delays. If regional databases do exist and are sufficiently accurate, regional comparisons could be made and built up to a global analysis.  

Finally, it would be worthwhile to conduct another full estimation of the total economic benefit generated as a result of GPS which includes aviation. Previous influential estimations have overlooked the benefits to the aviation sector. However, this paper illustrates that the aviation industry has been a beneficiary of GPS-wrought economic gains. Acquiring updated and more inclusive estimations of GPS’s economic benefit will contribute to policy conversations over its improvement. 

\section{Conclusion}
Global public goods, by their nature, are enjoyed by most people on earth, each of whom has an individual value for the benefit they receive. As such, small improvements to global public goods that increase the value they provide can multiply into large net welfare increases. In this paper, I focused on one such improvement that increased the accuracy of the location signals provided by GPS. Using a difference-in-differences estimator, I measured the impact of this change on delays gained in-flight, relative to the year before. I found that additional in-flight delays decreased by 0.617 minutes on average. I then stratified the flights by distance and number of months post change in order to test different hypotheses about how the change affected delays. This led to two additional findings, that delays decreased more on longer flights than shorter ones, relative to 1999, and that the effect of the change was initially growing but started to fall after a few months to the 0.617 minutes per flight level. In order to quantify the monetary value of these reduced delays, I calculated how much each passenger valued the time they saved and scaled that by the number of passengers each month. In total, I found that the GPS accuracy improvement led to a \$268 million net welfare increase in the first year. This value is over twice what the US appropriated to spend on the GPS system in the same year. 

These results are important because they fill a hole in the GPS valuation literature and quantifiably demonstrate how global public goods’ diffuse nature allows small changes to have large welfare impacts. Previous estimates of the economic benefit of GPS have both overlooked its impact on aviation and have not focused on the immediate benefits that stemmed from the decision to remove Selective Availability. My results demonstrate that GPS has indeed had a sizable economic impact on the aviation industry. Additionally, the size of the economic benefit I found, \$268 million, can help economists and policy makers understand the impact of a discrete global public good improvement. This is especially relevant in the context of current disruptive threats to GPS provision, namely jamming and spoofing. I discuss how usage of GPS jamming and spoofing tools can quickly lead to large welfare losses and provide an initial quantification of this loss. 

In order to mitigate the welfare losses from GPS jamming and spoofing, I propose that the US government invest in ways to guarantee accurate GPS signals. I give an overview of two of the initial proposals -- a ban on the use of GPS jamming and spoofing and the Space Force’s “Resilient GPS” program -- and discuss their relative strengths and weaknesses. In light of states’ demonstrated unwillingness to refrain from GPS jamming and spoofing, I argue that the US government should provision a physical upgrade to the GPS system, the “Resilient GPS” program or otherwise. I also caution against reimposing a policy similar to Selective Availability that sees these upgrades only accessible to the military.

Ultimately, the diffuse nature of welfare gains from global public good improvements also applies to welfare losses from disruptions to these same public goods. In order to generate large amounts of welfare, and protect existing welfare, policy makers should be willing to invest into global public goods, be they GPS, multinational scientific research, or collaborative security efforts.

\newpage
\printbibliography

\newpage
\appendix

\section{Estimated Welfare Gains Under Each Regression Scenario}
\label{sec:welfaretable}
\begin{table}[h!]
\centering
\caption{Welfare Estimates by Regression Group}
\label{tab:welfareestimates}
\begin{tabular}{lrrr}
\toprule
\textbf{Regression Group} & \textbf{Lower Bound} & \textbf{Valuation} & \textbf{Upper Bound} \\
\midrule
Full Year Regression & \$152,672,940 & \$183,464,122 & \$228,367,928 \\
\addlinespace
Subdivided by Recency & \$206,012,482 & \$247,561,218 & \$308,153,124 \\
\addlinespace

Subdivided by Distance & \$169,554,153 & \$203,749,949 & \$253,618,817 \\
\addlinespace

Subdivided by Recency and Distance & \$223,124,113 & \$268,123,934 & \$333,748,673 \\
\addlinespace

\bottomrule
\end{tabular}
\end{table}
\FloatBarrier

\section{Data Cleaning Overview}
\label{sec:datacleaning}
Unfortunately, the BTS on time performance data is not perfect. An audit of the data conducted by the US Department of Transportation Office of Inspector General concluded that the dataset may contain inaccuracies \parencite{smith_bureau_2024}. The authors wrote that “BTS relies on informal procedures to verify the accuracy of on-time performance data and lacks procedures to verify the data’s completeness” \parencite{smith_bureau_2024}. To address this potential source of inaccuracy, I removed observations from the dataset that were likely inaccurate.

I started with flights with in-air times less than 0. There were only 7 flights with negative air times. These negative flight times are most likely an error in reporting and due to the small number of affected flights relative to the size of the dataset, it would not affect the results of the study to remove them. Without the removed flights, the minimum arrival delay changed to -989. This value appears illogical, however the departure delay for that flight was -990 which means the flight was probably moved up relative to its scheduled time leaving 16.5 hours earlier than intended and landing 16.48 hours earlier than intended.

Additional discrepancies showed up in the $adddelay$ variable. The original minimum value of -1462 suggested that at least one flight in 1999 or 2000 made up 24 hours in flight. Considering that the longest domestic flights in the US are about 6 hours long, this minimum value is most likely a reporting error. There are 574 flights out of the over 11 million in which the flight gained over 1000 minutes from its departure delay to arrival delay. Adding to the evidence of a reporting error, 571 of these flights were operated by Delta Airlines and the other three were operated by American Airlines. Thus, I removed these 574 flights from the dataset. Additionally, I removed any flight where the absolute value of the $adddelay$ was greater than the actual elapsed time of the flight. This is because it is impossible for a flight to make up time on its delay or lose time on its delay in excess of the time it was away from the airports. For example, a 50 minute long flight would have to take 0 minutes from pulling away from its departure gate to pulling into its arrival gate to completely erase a 50 minute delay at the departing airport. On the other end, if a flight leaves on time and takes 50 minutes to arrive, it cannot be more than 50 minutes delayed on arrival. There were 624 flights in the dataset, including the 574 over 1000 minutes, that I removed for having an absolute value of their $adddelay$ larger than their actual elapsed time.

\section{Estimation Coefficients}
\subsection{Base Regression}
\noindent \emph{Calculated Value of Removing SA: \emph{\$183,464,122}}

\begin{table}[h!]
\centering
\caption{Estimation Coefficients of Base Regression}
\label{tab:escoefsbase}

\begin{tabularx}{\textwidth}{L{3cm} Y Y Y Y Y Y}
\toprule
\textbf{Month} & \textbf{0-500 Miles} & \textbf{500-1000 Miles} & \textbf{1000-1500 Miles} & \textbf{1500-2000 Miles} & \textbf{2000-2500 Miles} & \textbf{2500+ Miles} \\
\midrule
May ‘00 & 0.617 & 0.617 & 0.617 & 0.617 & 0.617 & 0.617 \\
June ‘00 & 0.617 & 0.617 & 0.617 & 0.617 & 0.617 & 0.617 \\
July ‘00 & 0.617 & 0.617 & 0.617 & 0.617 & 0.617 & 0.617 \\
August ‘00 & 0.617 & 0.617 & 0.617 & 0.617 & 0.617 & 0.617 \\
September ‘00 & 0.617 & 0.617 & 0.617 & 0.617 & 0.617 & 0.617 \\
October ‘00 & 0.617 & 0.617 & 0.617 & 0.617 & 0.617 & 0.617 \\
November ‘00 & 0.617 & 0.617 & 0.617 & 0.617 & 0.617 & 0.617 \\
December ‘00 & 0.617 & 0.617 & 0.617 & 0.617 & 0.617 & 0.617 \\
January ‘01 & 0.617 & 0.617 & 0.617 & 0.617 & 0.617 & 0.617 \\
February ‘01 & 0.617 & 0.617 & 0.617 & 0.617 & 0.617 & 0.617 \\
March ‘01 & 0.617 & 0.617 & 0.617 & 0.617 & 0.617 & 0.617 \\
April ‘01 & 0.617 & 0.617 & 0.617 & 0.617 & 0.617 & 0.617 \\

\bottomrule
\end{tabularx}
\end{table}
\FloatBarrier

\subsection{Regression by Month Groups}
\noindent \emph{Calculated Value of Removing SA: \emph{\$247,561,218}}
\begin{table}[h!]
\centering
\caption{Estimation Coefficients Using Month Based Regressions}
\label{tab:escoefsmonth}

\begin{tabularx}{\textwidth}{L{3cm} Y Y Y Y Y Y}
\toprule
\textbf{Month} & \textbf{0-500 Miles} & \textbf{500-1000 Miles} & \textbf{1000-1500 Miles} & \textbf{1500-2000 Miles} & \textbf{2000-2500 Miles} & \textbf{2500+ Miles} \\
\midrule
May ‘00 & 0.434 & 0.434 & 0.434 & 0.434 & 0.434 & 0.434 \\
June ‘00 & 1.508 & 1.508 & 1.508 & 1.508 & 1.508 & 1.508 \\
July ‘00 & 1.687 & 1.687 & 1.687 & 1.687 & 1.687 & 1.687 \\
August ‘00 & 1.185 & 1.185 & 1.185 & 1.185 & 1.185 & 1.185 \\
September ‘00 & 0.617 & 0.617 & 0.617 & 0.617 & 0.617 & 0.617 \\
October ‘00 & 0.617 & 0.617 & 0.617 & 0.617 & 0.617 & 0.617 \\
November ‘00 & 0.617 & 0.617 & 0.617 & 0.617 & 0.617 & 0.617 \\
December ‘00 & 0.617 & 0.617 & 0.617 & 0.617 & 0.617 & 0.617 \\
January ‘01 & 0.617 & 0.617 & 0.617 & 0.617 & 0.617 & 0.617 \\
February ‘01 & 0.617 & 0.617 & 0.617 & 0.617 & 0.617 & 0.617 \\
March ‘01 & 0.617 & 0.617 & 0.617 & 0.617 & 0.617 & 0.617 \\
April ‘01 & 0.617 & 0.617 & 0.617 & 0.617 & 0.617 & 0.617 \\

\bottomrule
\end{tabularx}
\end{table}
\FloatBarrier

\subsection{Regression by Distance Groups}
\noindent \emph{Calculated Value of Removing SA: \emph{\$203,749,949}}

\begin{table}[h!]
\centering
\caption{Estimation Coefficients Using Distance Based Regressions}
\label{tab:escoefsdistance}

\begin{tabularx}{\textwidth}{L{3cm} Y Y Y Y Y Y}
\toprule
\textbf{Month} & \textbf{0-500 Miles} & \textbf{500-1000 Miles} & \textbf{1000-1500 Miles} & \textbf{1500-2000 Miles} & \textbf{2000-2500 Miles} & \textbf{2500+ Miles} \\
\midrule
May ‘00 & 0.554 & 0.496 & 0.699 & 0.457 & 2.883 & 2.408 \\
June ‘00 & 0.554 & 0.496 & 0.699 & 0.457 & 2.883 & 2.408 \\
July ‘00 & 0.554 & 0.496 & 0.699 & 0.457 & 2.883 & 2.408 \\
August ‘00 & 0.554 & 0.496 & 0.699 & 0.457 & 2.883 & 2.408 \\
September ‘00 & 0.554 & 0.496 & 0.699 & 0.457 & 2.883 & 2.408 \\
October ‘00 & 0.554 & 0.496 & 0.699 & 0.457 & 2.883 & 2.408 \\
November ‘00 & 0.554 & 0.496 & 0.699 & 0.457 & 2.883 & 2.408 \\
December ‘00 & 0.554 & 0.496 & 0.699 & 0.457 & 2.883 & 2.408 \\
January ‘01 & 0.554 & 0.496 & 0.699 & 0.457 & 2.883 & 2.408 \\
February ‘01 & 0.554 & 0.496 & 0.699 & 0.457 & 2.883 & 2.408 \\
March ‘01 & 0.554 & 0.496 & 0.699 & 0.457 & 2.883 & 2.408 \\
April ‘01 & 0.554 & 0.496 & 0.699 & 0.457 & 2.883 & 2.408 \\

\bottomrule
\end{tabularx}
\end{table}
\FloatBarrier

\subsection{Regression by Distance Groups Within Month Groups}
\label{sec:regdistrecency}
\noindent \emph{Calculated Value of Removing SA: \emph{\$268,123,934}}

\begin{table}[h!]
\centering
\caption{Estimation Coefficients Using Regressions Subdivided by Distance and Month}
\label{tab:escoefsdistancemonth}

\begin{tabularx}{\textwidth}{L{3cm} Y Y Y Y Y Y}
\toprule
\textbf{Month} & \textbf{0-500 Miles} & \textbf{500-1000 Miles} & \textbf{1000-1500 Miles} & \textbf{1500-2000 Miles} & \textbf{2000-2500 Miles} & \textbf{2500+ Miles} \\
\midrule
May ‘00 & 0.315 & 0.179 & 1.392 & 0.434\footnote{The output of this regression was not significant so I substituted in the coefficient from the 2 month regression.} & 1.361 & 2.453 \\
June ‘00 & 1.180 & 1.490 & 2.551 & 1.836 & 2.154 & 1.806 \\
July ‘00 & 1.334 & 1.677 & 2.774 & 1.780 & 3.031 & 2.525 \\
August ‘00 & 0.959 & 1.064 & 2.051 & 1.215 & 2.772 & 2.223 \\
September ‘00 & 0.554 & 0.496 & 0.699 & 0.457 & 2.883 & 2.408 \\
October ‘00 & 0.554 & 0.496 & 0.699 & 0.457 & 2.883 & 2.408 \\
November ‘00 & 0.554 & 0.496 & 0.699 & 0.457 & 2.883 & 2.408 \\
December ‘00 & 0.554 & 0.496 & 0.699 & 0.457 & 2.883 & 2.408 \\
January ‘01 & 0.554 & 0.496 & 0.699 & 0.457 & 2.883 & 2.408 \\
February ‘01 & 0.554 & 0.496 & 0.699 & 0.457 & 2.883 & 2.408 \\
March ‘01 & 0.554 & 0.496 & 0.699 & 0.457 & 2.883 & 2.408 \\
April ‘01 & 0.554 & 0.496 & 0.699 & 0.457 & 2.883 & 2.408 \\

\bottomrule
\end{tabularx}
\end{table}
\FloatBarrier

\section{Regression Results by Flight Distance Within Recency-Based Subdivisions}

\subsection{1 Month Regression}

\begin{table}[h!]
    \centering
    \caption{1 Month Regression with Data Subdivided by Flight Distance}
    \label{tab:1monthdistancereg}
\begin{tabular}{lcccccc} \hline
 & 0--500 & 500--1000 & 1000--1500 & 1500--2000 & 2000--2500 & 2500+ \\
VARIABLES & adddelay & adddelay & adddelay & adddelay & adddelay & adddelay \\ \hline
 &  &  &  &  &  &  \\
treated & -0.689*** & -1.269*** & -1.617*** & -1.479*** & 0.447** & 1.334*** \\
 & (0.0348) & (0.0507) & (0.0918) & (0.138) & (0.217) & (0.380) \\
\addlinespace
GPS & -0.293*** & -0.647*** & 0.232** & -0.425*** & -0.858*** & -0.407 \\
 & (0.0370) & (0.0550) & (0.0989) & (0.144) & (0.221) & (0.384) \\
\addlinespace
treated\#GPS & -0.315*** & -0.179** & -1.392*** & -0.0614 & -1.361*** & -2.453*** \\
 & (0.0523) & (0.0774) & (0.139) & (0.203) & (0.317) & (0.567) \\
\addlinespace
destcanper & 0.385*** & 0.654*** & 0.645*** & 0.513*** & 0.528*** & 0.948*** \\
 & (0.00826) & (0.0117) & (0.0177) & (0.0228) & (0.0364) & (0.0549) \\
\addlinespace
origincanper & 0.336*** & 0.506*** & 0.625*** & 0.609*** & 0.885*** & 0.417*** \\
 & (0.00989) & (0.0137) & (0.0232) & (0.0328) & (0.0628) & (0.0719) \\
\addlinespace
Constant & -1.205*** & -2.379*** & -3.441*** & -3.700*** & -5.019*** & -5.985*** \\
 & (0.0289) & (0.0429) & (0.0752) & (0.123) & (0.209) & (0.336) \\
\addlinespace
Observations & 760,900 & 596,519 & 240,228 & 119,742 & 61,562 & 22,209 \\
R-squared & 0.031 & 0.050 & 0.043 & 0.031 & 0.030 & 0.038 \\

\hline
\multicolumn{7}{c}{Robust standard errors in parentheses} \\
\multicolumn{7}{c}{*** p$<$0.01, ** p$<$0.05, * p$<$0.1} \\
\end{tabular}
\end{table}
\FloatBarrier

\subsection{2 Month Regression}

\begin{table}[h!]
    \centering
    \caption{2 Month Regression with Data Subdivided by Flight Distance}
    \label{tab:2monthdistancereg}
\begin{tabular}{lcccccc} \hline
 & 0--500 & 500--1000 & 1000--1500 & 1500--2000 & 2000--2500 & 2500+ \\
VARIABLES & adddelay & adddelay & adddelay & adddelay & adddelay & adddelay \\ \hline
 &  &  &  &  &  &  \\
treated & 0.0516** & 0.0226 & -0.0887 & -0.296*** & 0.836*** & 1.527*** \\
 & (0.0238) & (0.0345) & (0.0622) & (0.0931) & (0.147) & (0.267) \\
\addlinespace
GPS & 0.745*** & 1.231*** & 2.248*** & 1.970*** & 2.184*** & 2.660*** \\
 & (0.0263) & (0.0390) & (0.0692) & (0.100) & (0.149) & (0.263) \\
\addlinespace
treated\#GPS & -1.180*** & -1.490*** & -2.551*** & -1.836*** & -2.154*** & -1.806*** \\
 & (0.0372) & (0.0553) & (0.0988) & (0.141) & (0.218) & (0.394) \\
\addlinespace
destcanper & 0.344*** & 0.599*** & 0.624*** & 0.477*** & 0.498*** & 0.799*** \\
 & (0.00531) & (0.00772) & (0.0117) & (0.0160) & (0.0234) & (0.0344) \\
\addlinespace
origincanper & 0.322*** & 0.463*** & 0.554*** & 0.683*** & 0.813*** & 0.429*** \\
 & (0.00601) & (0.00854) & (0.0146) & (0.0195) & (0.0348) & (0.0437) \\
\addlinespace
Constant & -1.753*** & -3.475*** & -4.548*** & -5.010*** & -6.264*** & -7.334*** \\
 & (0.0198) & (0.0299) & (0.0531) & (0.0838) & (0.142) & (0.241) \\
\addlinespace
Observations & 1,529,640 & 1,197,710 & 483,908 & 241,631 & 124,142 & 45,597 \\
R-squared & 0.029 & 0.044 & 0.037 & 0.038 & 0.029 & 0.033 \\

\hline
\multicolumn{7}{c}{Robust standard errors in parentheses} \\
\multicolumn{7}{c}{*** p$<$0.01, ** p$<$0.05, * p$<$0.1} \\
\end{tabular}
\end{table}
\FloatBarrier

\subsection{3 Month Regression}

\begin{table}[h!]
    \centering
    \caption{3 Month Regression with Data Subdivided by Flight Distance}    \label{tab:3monthdistancereg}
\begin{tabular}{lcccccc} \hline
 & 0--500 & 500--1000 & 1000--1500 & 1500--2000 & 2000--2500 & 2500+ \\
VARIABLES & adddelay & adddelay & adddelay & adddelay & adddelay & adddelay \\ \hline
 &  &  &  &  &  &  \\
treated & 0.329*** & 0.436*** & 0.679*** & 0.286*** & 1.779*** & 2.778*** \\
 & (0.0194) & (0.0282) & (0.0510) & (0.0767) & (0.121) & (0.222) \\
\addlinespace
GPS & 1.012*** & 1.864*** & 2.965*** & 2.384*** & 3.462*** & 4.122*** \\
 & (0.0216) & (0.0321) & (0.0560) & (0.0807) & (0.124) & (0.218) \\
\addlinespace
treated\#GPS & -1.334*** & -1.677*** & -2.774*** & -1.780*** & -3.031*** & -2.525*** \\
 & (0.0301) & (0.0446) & (0.0783) & (0.112) & (0.176) & (0.314) \\
\addlinespace
destcanper & 0.332*** & 0.589*** & 0.531*** & 0.463*** & 0.508*** & 0.792*** \\
 & (0.00399) & (0.00593) & (0.00872) & (0.0118) & (0.0171) & (0.0249) \\
\addlinespace
origincanper & 0.279*** & 0.403*** & 0.417*** & 0.476*** & 0.557*** & 0.242*** \\
 & (0.00431) & (0.00635) & (0.0104) & (0.0138) & (0.0241) & (0.0299) \\
\addlinespace
Constant & -1.971*** & -4.033*** & -5.059*** & -5.302*** & -6.911*** & -8.474*** \\
 & (0.0163) & (0.0255) & (0.0442) & (0.0688) & (0.117) & (0.202) \\
\addlinespace
Observations & 2,269,281 & 1,775,939 & 717,964 & 360,269 & 185,142 & 68,479 \\
R-squared & 0.027 & 0.042 & 0.030 & 0.028 & 0.024 & 0.035 \\

\hline
\multicolumn{7}{c}{Robust standard errors in parentheses} \\
\multicolumn{7}{c}{*** p$<$0.01, ** p$<$0.05, * p$<$0.1} \\
\end{tabular}
\end{table}
\FloatBarrier

\subsection{4 Month Regression}

\begin{table}[h!]
    \centering
    \caption{4 Month Regression with Data Subdivided by Flight Distance}    \label{tab:4monthdistancereg}
\begin{tabular}{lcccccc} \hline
 & 0--500 & 500--1000 & 1000--1500 & 1500--2000 & 2000--2500 & 2500+ \\
VARIABLES & adddelay & adddelay & adddelay & adddelay & adddelay & adddelay \\ \hline
 &  &  &  &  &  &  \\
treated & 0.224*** & 0.304*** & 0.588*** & 0.158** & 1.754*** & 2.589*** \\
 & (0.0174) & (0.0251) & (0.0450) & (0.0695) & (0.108) & (0.200) \\
\addlinespace
GPS & 1.000*** & 1.879*** & 2.982*** & 2.319*** & 3.762*** & 4.415*** \\
 & (0.0188) & (0.0278) & (0.0482) & (0.0703) & (0.109) & (0.194) \\
\addlinespace
treated\#GPS & -0.959*** & -1.064*** & -2.051*** & -1.215*** & -2.772*** & -2.223*** \\
 & (0.0260) & (0.0384) & (0.0672) & (0.0976) & (0.154) & (0.275) \\
\addlinespace
destcanper & 0.306*** & 0.495*** & 0.455*** & 0.428*** & 0.466*** & 0.755*** \\
 & (0.00327) & (0.00496) & (0.00790) & (0.0114) & (0.0140) & (0.0223) \\
\addlinespace
origincanper & 0.264*** & 0.353*** & 0.387*** & 0.432*** & 0.513*** & 0.219*** \\
 & (0.00341) & (0.00497) & (0.00816) & (0.0114) & (0.0188) & (0.0245) \\
\addlinespace
Constant & -1.955*** & -3.903*** & -5.160*** & -5.347*** & -7.079*** & -8.994*** \\
 & (0.0148) & (0.0228) & (0.0404) & (0.0644) & (0.104) & (0.184) \\
\addlinespace
Observations & 3,031,899 & 2,371,759 & 960,911 & 483,413 & 248,170 & 92,438 \\
R-squared & 0.030 & 0.038 & 0.028 & 0.026 & 0.024 & 0.035 \\

\hline
\multicolumn{7}{c}{Robust standard errors in parentheses} \\
\multicolumn{7}{c}{*** p$<$0.01, ** p$<$0.05, * p$<$0.1} \\
\end{tabular}
\end{table}
\FloatBarrier

\section{AI Acknowledgment}
I’d like to acknowledge my use of generative AI in this project. Specifically, I used ChatGPT to aid with Stata and LaTex coding, finding data sources, and for minor writing edits. None of the data used in this project was downloaded directly from generative AI. 

\begin{center}
    \noindent \textit{Fin}
\end{center}

\end{document}